\documentclass{JHEP3}
\pdfoutput=1
\usepackage[pdftex]{graphicx}
\usepackage{mathrsfs}
\usepackage{amsmath, amsthm, amssymb}
\usepackage{verbatim}

\newcommand{\be}{\begin{equation}}
\newcommand{\ee}{\end{equation}}
\newcommand{\bea}{\begin{eqnarray}}
\newcommand{\eea}{\end{eqnarray}}

\newcommand{\mc}{\mathcal}

\title{New techniques for chargino-neutralino detection at LHC }

\author{Maria Eugenia Cabrera\\
        University of Amsterdam\\
        Institute of Theoretical Physics\\
        GRAPPA\\
        E-mail: \email{M.E.CabreraCatalan@uva.nl}}

\author{J. Alberto Casas \\
        Instituto de F\'isica Te\'orica, IFT-UAM/CSIC \\
        U.A.M., Cantoblanco, \\
        28049 Madrid, Spain \\
        E-mail: \email{alberto.casas@uam.es}} 

\author{Bryan Zald\'ivar \\
        Instituto de F\'isica Te\'orica, IFT-UAM/CSIC \\
        U.A.M., Cantoblanco, \\
        28049 Madrid, Spain \\
        E-mail: \email{b.zaldivar.m@csic.es}}

\abstract{\small
The recent LHC discovery of a Higgs-like boson at 126 GeV has important consequences for SUSY, pushing the spectrum of strong-interacting supersymmetric particles to high energies, very difficult to probe at the LHC. This gives extra motivation to study 
the direct production of electroweak particles, as charginos and neutralinos, which are presently very poorly constrained. The aim of this work is to improve the analysis of chargino-neutralino pair production at LHC, focusing on the kinematics of the processes. We propose a new method based on the study of the {\it poles} of a certain kinematical variable. This complements other approaches, giving new information about the spectrum and improving the signal-to-background ratio. We illustrate the method in particular SUSY models, and show that working with the LHC at 100/fb luminosity one would be able to distinguish the SUSY signal from the Standard Model background.}

\keywords{Beyond Standard Model, Collider physics, Supersymmetry
  Searches, Kinematic Variables, LHC physics, chargino neutralino production}
\preprint{IFT-UAM/CSIC-12-114}

\begin{document}

\section{Introduction}

%Very recently, the LHC seems to have done a discovery that scientific community has waited for decades. Indeed, it is quite probable that the new discovered particle is the Standard Model (SM) Higgs boson, with a mass around 126 GeV \cite{HiggsATLAS}, \cite{HiggsCMS}. 

%On the other hand, Supersymmetry (SUSY) is probably the best motivated (and more extensively studied) beyond SM scenario, both from the formal as well as from the phenomenological point of view, giving detailed predictions of New Physics (NP) signals.  Indeed, it has became a paradigmatic framework for experimentalists's analysis of constraints on NP when studying LHC data. 

The recent discovery of the Higgs boson, with a mass around 126 GeV \cite{HiggsATLAS}, \cite{HiggsCMS}, does not only have crucial importance by itself. It also has far-reaching consequences for well-motivated candidates for physics beyond the Standard Model (SM), such as Supersymmetry (SUSY), and in particular the Minimal Supersymmetric Standard Model (MSSM). As it is well known, such a value requires large radiative
corrections, which go with (the logarithm of) the supersymmetric masses, in particular with the stop masses. Consequently, the latter
must be rather high (well above 1 TeV unless the stop mixing is close to the maximal value), thus suggesting that the mass scale of SUSY particles could be substantially higher than expected from fine-tuning arguments. This would also make very challenging, if not impossible, to detect SUSY at LHC in a direct or indirect way. In fact, this is already the most likely situation for the constrained MSSM~\cite{Farina:2011bh,Balazs:2012qc,Akula:2012kk,Buchmueller:2012hv,Arbey:2012dq,Strege:2012bt,Cabrera:2012vu}, i.e.\ assuming universality of soft terms at a high-scale.

Prospects become much more interesting if some supersymmetric states remain sufficiently light, which in general implies to go beyond the constrained MSSM. An attractive possibility in this sense is that charginos and neutralinos are substantially lighter than sfermions. This scenario is supported not only by the phenomenological fact that the present bounds on charginos and neutralinos are pretty mild. There are also some theoretical and phenomenological motivations to explore this possibility. Namely, the successful supersymmetric unification of the gauge couplings requires light supersymmetric fermions. Besides, heavy {{{sfermions}}} are welcome to suppress dangerous flavor-violation effects. Another motivation comes from dark matter (DM) constraints. The last data of XENON100 in combination with the Higgs mass have narrowed enormously the MSSM candidates for DM, see e.g. refs.\cite{Farina:2011bh, Cabrera:2012vu}. Probably the most satisfactory scenario that survives occurs when 
 the lightest neutralino is almost pure Higgsino with a mass around 1 TeV. But if gaugino masses are not universal there are other possibilities. Namely, one can have a much lighter (mostly bino) neutralino which annihilates in the early universe through a combination of $Z-$boson and Higgs funnels, see \cite{Farina:2011bh}. It can also co-annihilate with e.g. sleptons if their masses are close enough. Another alternative is that the LSP neutralino is mostly wino, in which case the co-annihilation with the lightest chargino becomes important.
 
In summary, a scenario where all the supersymmetric particles are too heavy, except charginos and neutralinos (and maybe gluinos), is plausible and has interesting motivations. Therefore it would be worthy to improve the present techniques to analyze the production and detection of chargino/neutralino pairs at the LHC; and this is the main motivation of this paper.

%For SUSY models with large hierarchy among their masses, the associated production of chargino-neutralino could be the best (and maybe unique) hope of detection. 
 
\begin{figure}[t]
\centering 
a)~\includegraphics[width=0.3\linewidth]{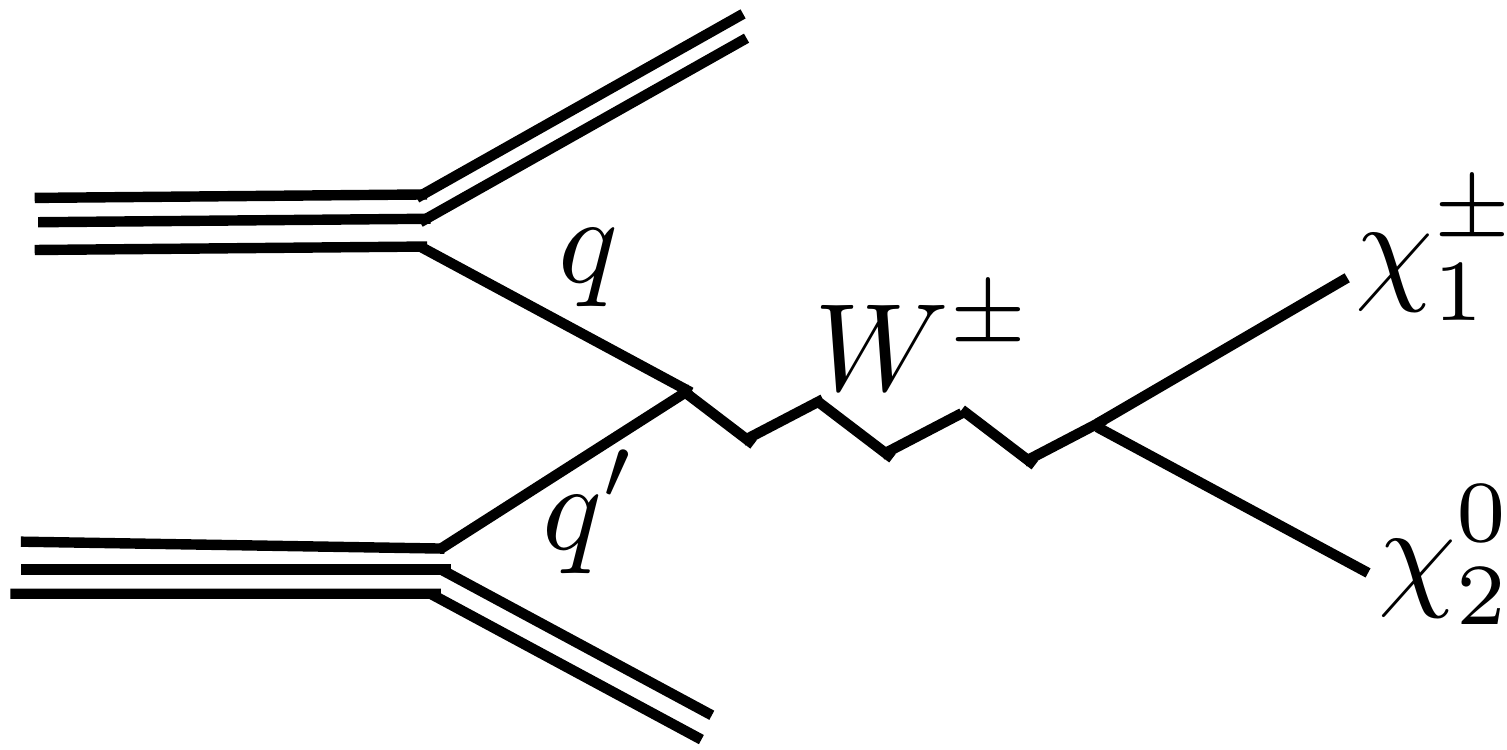}
\hspace{1cm}
b)~\includegraphics[width=0.25\linewidth]{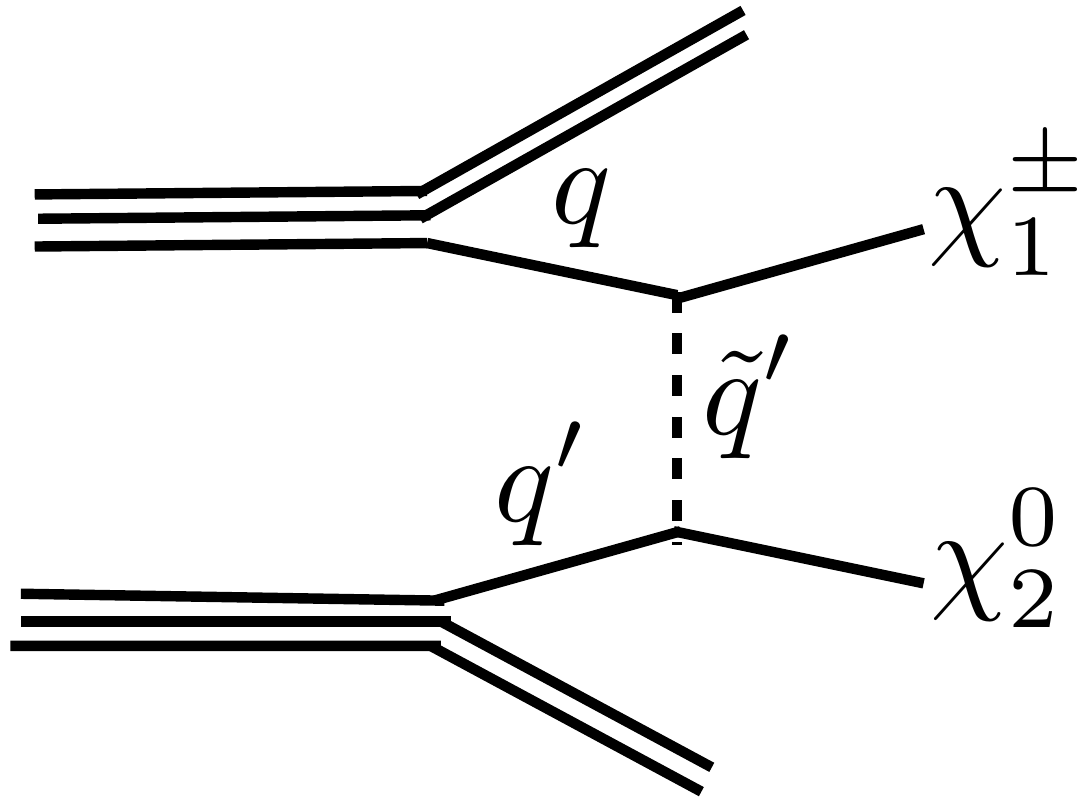}
\hspace{1cm}
c)~\includegraphics[width=0.3\linewidth]{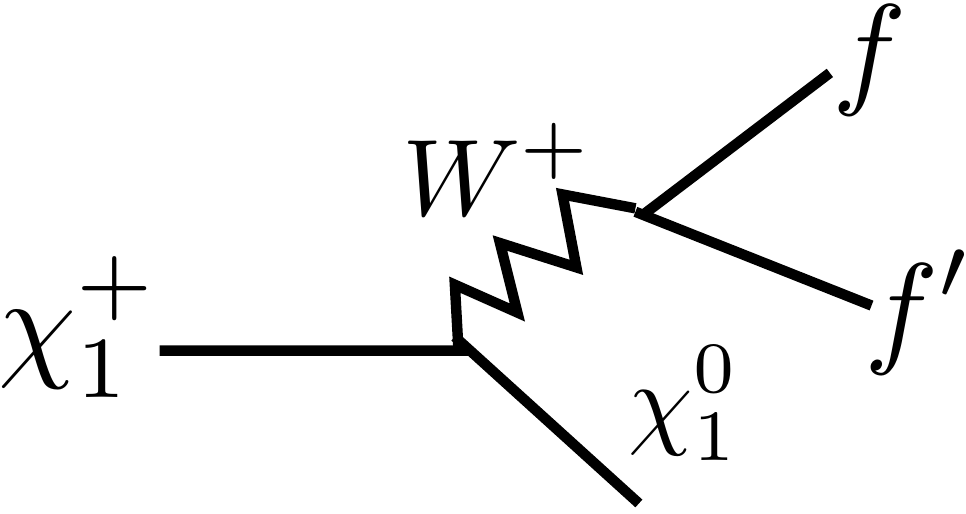}
\hspace{1cm}
d)~\includegraphics[width=0.3\linewidth]{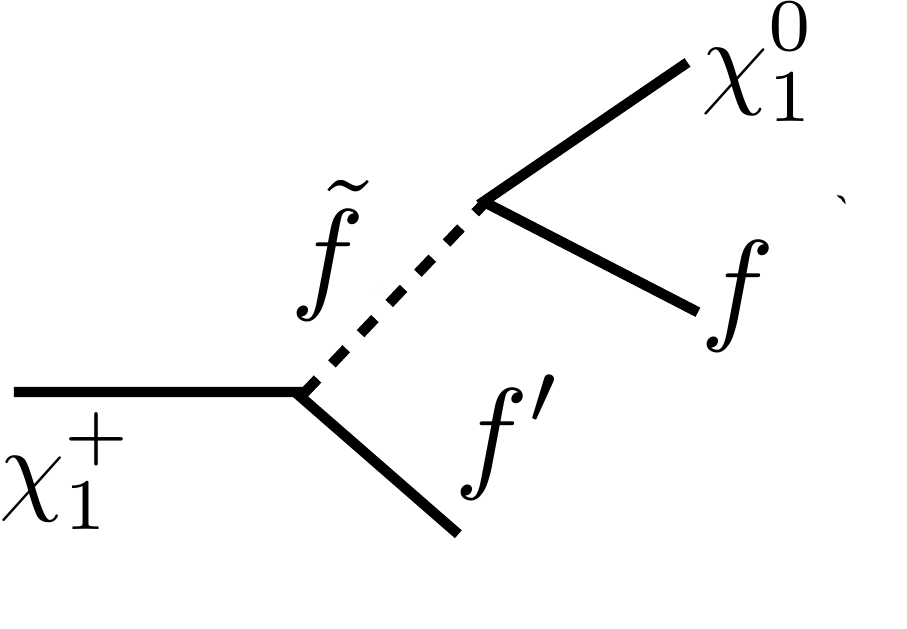}
\hspace{1cm}
e)~\includegraphics[width=0.3\linewidth]{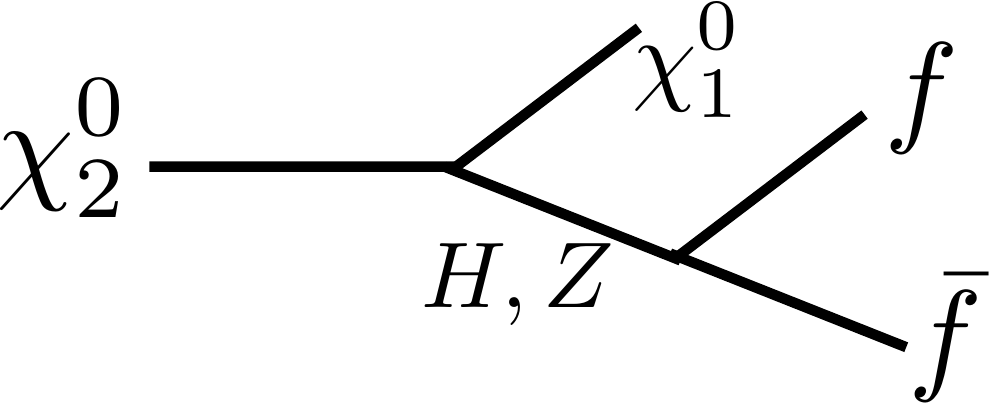}
\hspace{1cm}
f)~\includegraphics[width=0.3\linewidth]{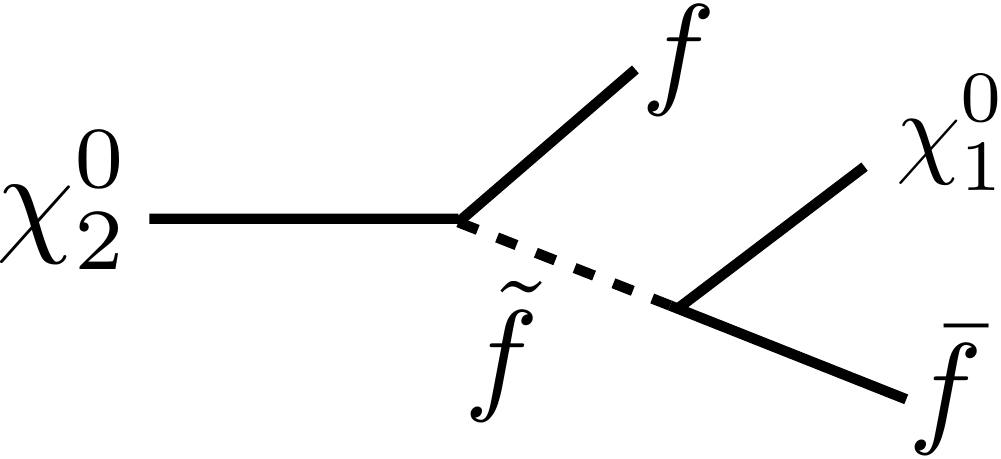}
\caption{\small{a) and b) Typical production processes of a pair  $\chi^{\pm}_1\chi^0_2$ in the LHC. c) and d) Chargino decay modes. e) and f) Neutralino decay modes.}} 
\label{fig:C1N2}
\end{figure}

In most of the cases the chargino-neutralino pair created is $\chi^{\pm}_1\chi^0_2$, i.e. the lightest chargino and the next-to-lightest neutralino. Some of the diagrams of production and decay are shown in Fig. \ref{fig:C1N2}.  The chargino and the neutralino can decay in several ways, always giving an LSP ($\chi^0_1$) at the end of each cascade. 

%%%%%%%%%%%%%%%%%%%%%%%%%%%%%%%%%%%%%%%%%%%%%%%%%%
%%%%%%%%%%%%%%%%%%%%%%%%%%%%%%%%%%%%%%%%%%%%%%%%%%

The study of chargino-neutralino pair production has been performed previously in e.g. \cite{Aad:2012cwa}, \cite{Chatrchyan:2012ewa}, \cite{Aad:2009wy} through the analysis of leptonic final states. The aim of our work is to improve those analyses, by proposing new strategies which are complementary and more efficient in some cases \footnote{For a general review on kinematical techniques, see \cite{Barr:2010zj}. }. Our analysis is by construction independent of the diagrams through which the chargino and the neutralino decay, since it is entirely based on the properties of the final states. It can also be applied to final states including hadrons, such as $b\bar b\ell$. 

The outline of the paper is as follows. In section \ref{sec:Art} we describe the present status of the analyses of possible chargino-neutralino production and detection, as done by ATLAS and CMS groups, and motivate the convenience of improving this kind of analysis.  In section \ref{sec:strat} we describe our strategy, by proposing a kinematical variable which allows to obtain direct information about the SUSY spectrum. It is also very useful to concentrate signal-events, thus improving the S/B ratio, and thus the LHC potential for discovery of new physics.
Section \ref{sec:Testing} is devoted to illustrate the use of this variable in concrete SUSY models. We finally conclude in section \ref{sec:Concl}.

\section{State of the art of the analysis}
\label{sec:Art}

As shown in Fig.\ref{fig:C1N2}, the neutralino $\chi^0_2$ can decay through a charged sfermion, a Higgs or a $Z-$boson, depending on the kinematical availability and the $\chi^0_2$ composition. Likewise, the chargino $\chi_1^\pm$ can decay, among other channels, through a charged sfermion or a $W-$boson. 
%If $\chi^\pm_1$ and $\chi^0_2$ decay through right-handed (left-handed) sleptons\footnote{$\nu_R$, $\tilde\nu_R$ are not considered in this analysis.},  the final state will contain $3\ell+\nu$ ($\ell+3\nu$), plus the two $\chi^0_1$ neutralinos.
However, as long as the sleptons are heavier than $\chi^0_2$ and $\chi^\pm_1$, these decay channels become very suppressed and the decays through on-shell Higgs, $Z$ and $W$ dominate.
\newline\newline
Ref. \cite{Aad:2012cwa} contains a study of purely leptonic final states of the type $\ell^+\ell^-\ell'^{\pm}$, where $\ell$ and $\ell'$ may be identical leptons. The authors perform separate analyses of two cases: 1) the invariant mass of identical opposite-sign leptons ($m_{\ell^+\ell^-}$) does {\em not} reproduce the $Z$-boson mass, $M_Z$, and 2) it does. They assume that the $\chi^0_2$ decays through a slepton (case 1) or a $Z-$boson (case 2).
In both cases, they use the $E_T^{\rm miss}$ variable to compare the actual experimental data with the SM background, finding no significant excess of events so far, once all the uncertainties are taken into account. The negative result is then interpreted as contour bounds in the parameter space of e.g. concrete simplified models.

A very complete analysis was presented later in ref.\cite{Chatrchyan:2012ewa}. Again, the authors focus on 3-lepton final states, but  using either the variable $E_T^{\rm miss}$ or $m_{\ell^+\ell^-}$ in combination with $M_T$ (the transverse mass built with the momentum of the unpaired lepton) and $E_T^{\rm miss}$. One option gives better sensitivity than the other, depending on the mass splittings between $\chi^0_1$ and its respective mothers. A second analysis was performed for 2-lepton final states, considering the possibility that one of the 3 leptons produced by $\chi^\pm_1$ or $\chi^0_2$ may be lost or does not pass the kinematical cuts. In a last analysis they considered that both $\chi^\pm_1$ and $\chi^0_2$ may decay through on-shell vector bosons, giving $2\ell+2j$ in the final state, for which the SM background has not intrinsic $E_T^{\rm miss}$. Using these techniques they were able to provide contour bounds on $\chi^\pm_1,\chi^0_2$ and $\chi^0_1$ masses for particular simplified models. 

A quite efficient analysis of the decay through sleptons was presented in ref. \cite{Aad:2009wy} (based on \cite{Nojiri:1999ki} and \cite{DeSanctis:2007td}), where the authors used the $m_{\ell^+\ell^-}$ variable. An obvious advantage of this choice us that  $m_{\ell^+\ell^-}$ is Lorentz invariant and thus the analysis is fully valid in the (boosted) LAB frame. It can be shown that, when the intermediate sleptons are produced on-shell, the histogram on $m_{\ell^+\ell^-}$ has an edge at a value given by a certain combination of the $\chi^0_2, \chi^0_1$ and $\tilde\ell$ masses\cite{Nojiri:1999ki},\cite{Gjelsten:2004ki},\cite{Birkedal:2005cm}. This potentially gives a very distinctive experimental signal. When the sleptons are produced off-shell, the authors use a strategy based on the study of end-points in $m_{\ell^+\ell^-}$. This case is much less efficient, since by definition one has poorer statistics and a more difficult-to-control background.

%Note that most of the previous analyses either do not provide direct information about the relevant SUSY spectrum or are quite inefficient due to poor statistics (with the exception of the $\chi_2^0-$decay through on-shell sleptons based on $m_{\ell^+\ell^-}$). In addition, they are designed to 
%work for specific channels of decay (through sleptons or $Z-$boson).

In this paper we propose the use of a variable whose distinctive feature is to concentrate the signal events around a peak, not relying on end-points. It also provides direct information about the spectrum. Besides, it can be applied without any assumption or guess about the decay mode that takes place (through $Z-$boson, sleptons, Higgs or whatever). We describe the idea in the next section.

\section{Our Strategy}
\label{sec:strat}

\subsection{The visible transverse energy}

In contrast with the previous analyses, our strategy is purely kinematical and based only on the characteristics of the initial ($\chi^0_2$) and the final states ($f\bar f\chi^0_1$); so it is independent of the channel through which the $\chi^0_2$ decays. As it is based just on the $\chi^0_2-$chain, the analysis can be easily combined with other analyses that use partial information from both the $\chi^0_2$ and the $\chi^\pm_1$ chains. It can also be applied to other processes where one or two $\chi^0_2$ states are produced.

We initially work in the reference frame in which  $\chi^0_2$ is at rest, which we call CM$\chi$ (do not confuse with the center-of mass of the partonic collision) and consider the ${\cal E}_T$ variable, defined as 
\be
\label{ET}
{\cal E}_T=\hat{E}_T^v + \hat{E}_T^\chi 
\ee
where $\hat{E}_T^v$, $ \hat{E}_T^\chi$ are  the transverse energies of the the visible system (e.g. $v\equiv \ell^+\ell^-$) and the missing system,
\be
\label{ETdef}
\hat{E}_T^v   =   \sqrt{M_v^2 + (\hat{p}^v_T)^2} ,\hspace{1cm} \hat{E}_T^\chi =  \sqrt{M_\chi^2 + (\hat{p}^\chi_T)^2} 
\ee
Here $M_v$ and $M_\chi$ are the invariant masses of the visible system and the LSP ($\chi_1^0$) and hats denote CM$\chi$ quantities everywhere. Of course, $(\hat{p}^v_T)^2=(\hat{p}^\chi_T)^2$.

As it has been discussed in ref. \cite{Cabrera:2012cj}, the histogram of events displayed in the ${\cal E}_T$ variable has a pole at the mass of $\chi_2^0$,
\be
\label{ETpole}
\left.{\cal E}_T\right|_{\rm pole}= E_{{\rm CM}\chi} = M_{\chi_2} 
\ee
Besides, for $\mc E_T>M_2$ there are no events, so the histogram has an sharp edge at the pole. Hence, potentially, the $\mc E_T$ variable can give a very distinctive signal, well separated from the background, providing in addition direct information about the SUSY spectrum. Note also that $\mc E_T$ is invariant under longitudinal boosts. There are however some problems. First, even working at the CM$\chi$ frame, i.e. assuming that $\chi^0_2$ was produced with vanishing transverse momentum,
we cannot measure the invisible transverse energy, $E_T^\chi$, due to the uncertainty on the value of $M_\chi$. Actually, we have checked that in general $M_\chi\simeq 0$ is not a good approximation. Second, the $\chi_2^0$ neutralino is usually produced with a non-vanishing transverse momentum. We postpone the second issue to the next subsection and focus now on the first one.

In order to avoid the dependence on unknown quantities, a good strategy is to work just with the visible transverse energy, $E^v_T$, defined in eq.(\ref{ETdef}). It can be easily checked that the pole in the $\mc E_T$ variable, eq.(\ref{ETpole}), translates into a pole in $\hat{E}^v_T$,
\be
\label{ETvpole}
\left. \hat{E}_T^v\right|_{\rm pole}= \frac{1}{2 M_{\chi_2}}\left[M_{\chi_2}^2 -M_\chi^2+M_v^2\right]
\ee
In general, the mass of the visible system, $M_v$, can change from event to event. However, if $\chi_2^0$ decays through a Higgs (very common case) or through a $Z-$boson, all the events are concentrated around $M_v = m_h$ ($M_Z$). An example of this kind, which shows clearly the edge and pole in the CM$\chi$ frame can be seen in the plot a) of Fig.2, to be discussed below in more detail.
When $\chi_2^0$ decays through sleptons, we can select a fraction of the events with similar $M_v$, and check that in CM$\chi$ a pole appears as in (\ref{ETvpole}). This decreases the statistics but not in a dramatic way. Actually, in cases where
$M_v^2 \ll M_{\chi_2}^2$, one could do a histogram with all the events, since the pole appears then around $\hat{E}_T^v\simeq \frac{1}{2M_{\chi_2}}(M_{\chi_2}^2-M_\chi^2)$. 

\noindent
Next, we study what happens when we go from the CM$\chi$ to the actual LAB frame.

\subsection{From CM$\chi$ to LAB}

\begin{figure}[t]
\centering 
\includegraphics[width=0.46\linewidth]{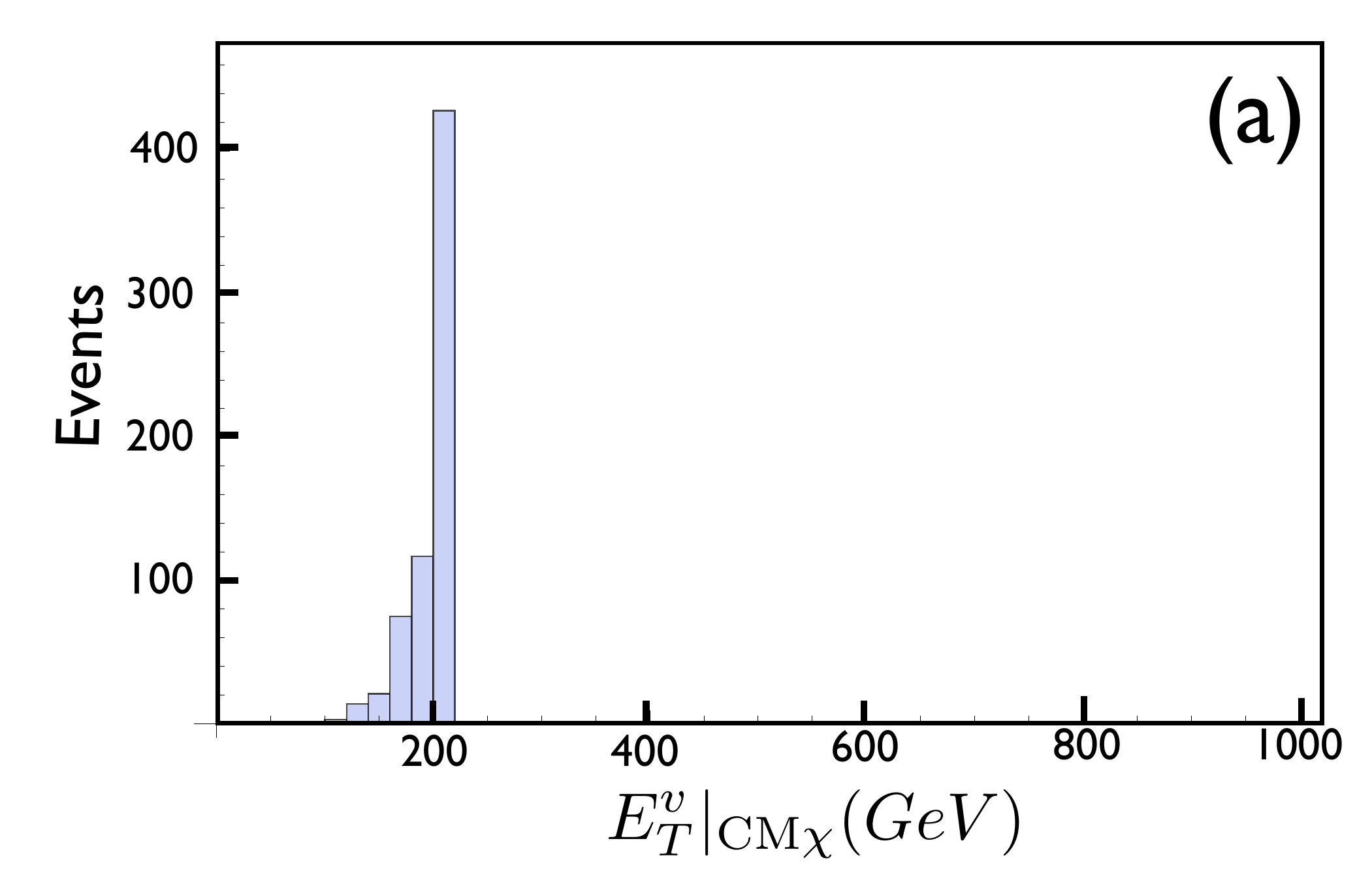}
\includegraphics[width=0.46\linewidth]{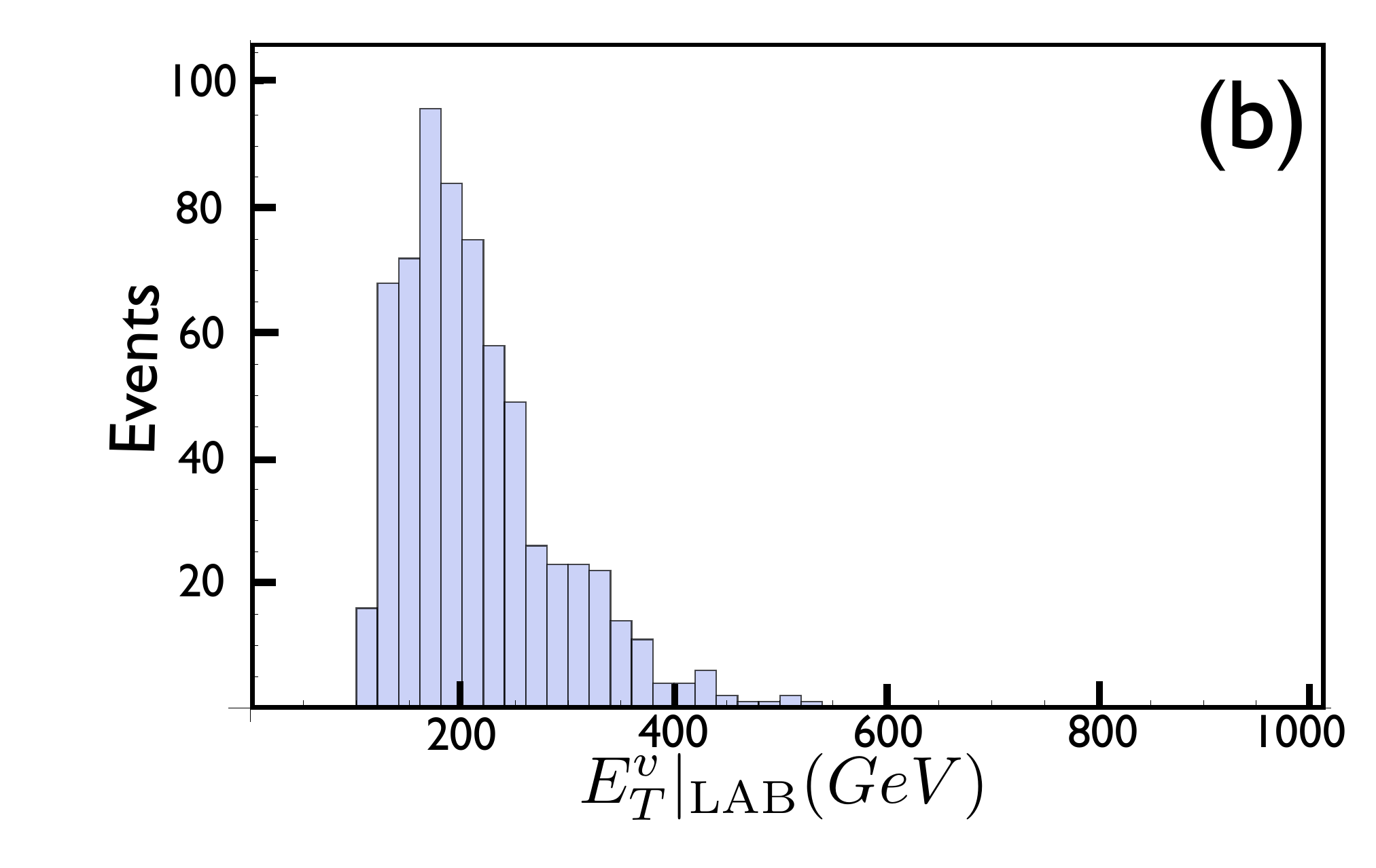}
\hspace{1cm}
\includegraphics[width=0.46\linewidth]{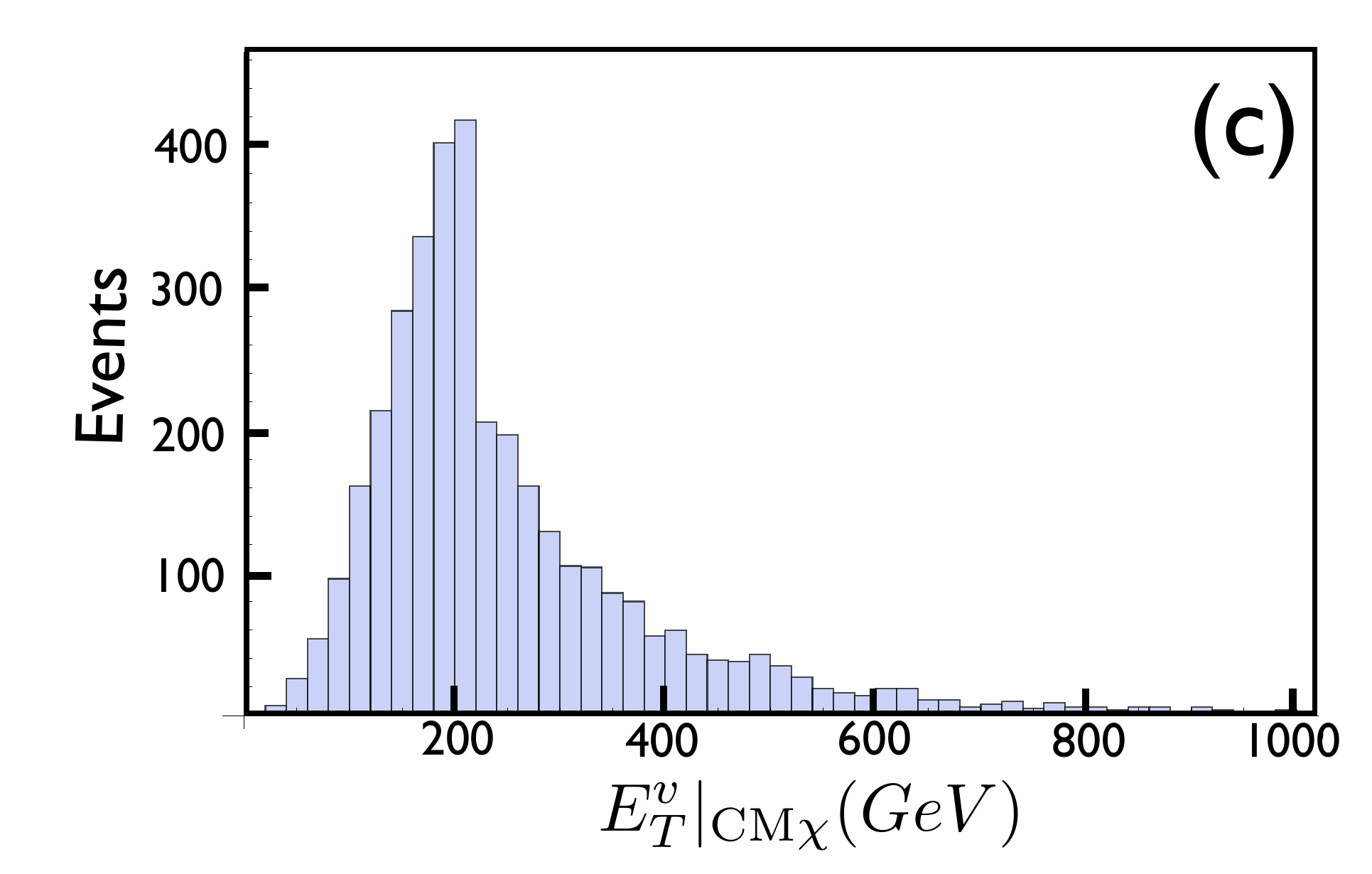}
\includegraphics[width=0.46\linewidth]{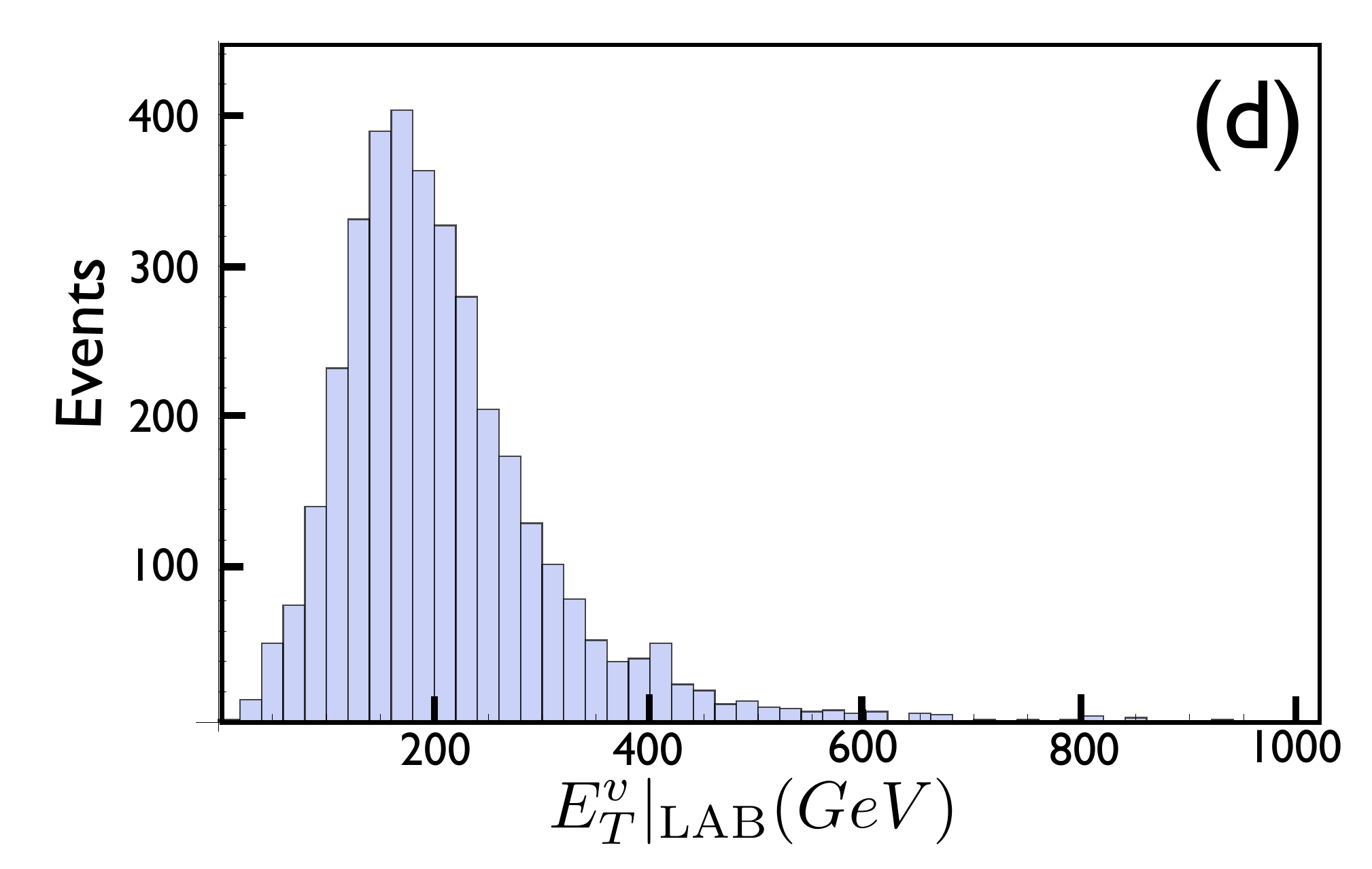}
\caption{\small{Histogram (number of events) of $E_T^v$ in the cases where $v$ is a) a $\tau\bar\tau$ pair in the CM$\chi$ frame; b) a $\tau\bar\tau$ pair in the LAB frame; c) a $j_bj_b$ pair in the CM$\chi$ frame, and d) a $j_bj_b$ pair in the LAB frame. The details of the SUSY point correspondent to this histograms are described in the text. }}
\label{fig:ETvCMvsLAB}
\end{figure}

The clean and sharp behavior of the $\hat{E}_T^v$  variable in CM$\chi$ becomes of course less keen-edged when passing to the LAB frame. This is illustrated by the plots of Fig. \ref{fig:ETvCMvsLAB}. 
%%%%%%%%%%%%%%%%%%%%%%%%%%%%%%%%%%%%%%%%%%%%%%%%%%
%%%%%%%%%%%%%%%%%%%%%%%%%%%%%%%%%%%%%%%%%%%%%%%%%%
They correspond to a CMSSM model with $m_0=500$ GeV, $M_{1/2}=700$ GeV, $\tan\beta=10$, $A_0=0$, and $\mu>0$. The associated relevant spectrum for us is $M_{\chi_2}\approx 554$ GeV and $M_\chi\approx294$ GeV. The $\chi_2^0$ neutralino decays almost entirely through an on-shell Higgs. In the CM$\chi$ frame the pole for $E_T^v$, eq. (\ref{ETvpole}), is at $\sim210$ GeV.  We have simulated the events using {\tt Pythia}, with a center-of-mass energy $\sqrt{s}=7$ TeV and a luminosity of 20/fb. The upper (lower) plots of Fig.2 correspond to events where the Higgs decayed into $\tau\bar\tau$ (a pair of b-jets, $j_b j_b$). The left (right) plots are obtained in the CM$\chi$ (LAB) frame.   We emphasize that this example is just for illustrative purposes. For this reason we have not incorporated a realistic $\tau$-reconstruction, as well as b-jet identification. 
%Note also that the Higgs mass is not the experimental one, but this is irrelevant at this point.

As expected, the CM$\chi$ histograms show a very clear and sharp peak (pole) around 200--220 GeV, followed by an edge.  In the case of the $j_bj_b$ histogram there appear events beyond the edge due to the limited efficiency in the jet reconstruction. The corresponding LAB histograms show two (non-dramatic) differences. First, a logical spreading around the maximum; and second, a slight shift of the peak towards smaller values of $E_T^v$. This shift is almost invisible for the $j_bj_b$ histogram. All these effects can be explained, estimated in a semi-analytical way and kept under control, as we discuss below.

We recall that the transverse variables we are using are only affected by transverse boosts when passing from CM$\chi$ to LAB. Longitudinal boosts are irrelevant for this analysis. There are two sources of transverse boosts. First, the partonic collision will not occur in general in its center-of-mass, i.e. it will have a non-vanishing net transverse momentum, typically due to initial state radiation. Second (and more importantly) even at the center-of-mass of the partonic collision, $\chi^0_2$ can be produced with non-vanishing transverse momentum (and opposite to that of $\chi_1^{\pm}$) . Next we estimate the change in the $E_T^v$ of the events at the pole due to the non-vanishing transverse momentum of $\chi_2^0$ in the LAB.

We will distinguish two perpendicular directions in the transverse plane: the direction along $p_T^{\chi_2}$ ($\parallel$), and the perpendicular to it ($\perp$).  Due to the boost in the $\parallel$ direction (we will ignore the boost in the longitudinal direction, which is irrelevant here), the visible energy changes (from CM$\chi$ to LAB) as
\be
\label{Ev}
\hat{E}^v\rightarrow E^v=\gamma\hat{E}^v - \beta\gamma\ (\hat{p}_{T}^{\ v})_\parallel
\ee
where  $(\hat{p}_{T}^{\ v})_\parallel$ is the component of the visible 3-momentum in the $\parallel$ direction. As usual, $\gamma$ and $\beta$ are the parameters of the Lorentz transformation, satisfying  
\be
\label{gammabeta}
\beta\gamma=\frac{\left|p_T^{\ \chi_2}\right|}{M_{\chi_2}} ~.
\ee

Now, for the events {\it at the pole} $\hat E^v_T = \hat E_v$ in the CM$\chi$ frame \cite{Cabrera:2012cj}. This holds after the transverse boost. Hence, when going to LAB, for those events the transverse energy of those events changes as in eq.(\ref{Ev}):
\be
\label{ETvLAB}
\hat{E}_T^v\rightarrow E_T^v=\gamma\hat{E}_T^v - \beta\gamma\ (\hat{p}_{T}^{\ v})_\parallel
\ee
which represents a shift
\be
\label{DeltaETv}
\Delta\hat{E}_T^v=(\gamma-1)\hat{E}_T^v - \beta\gamma\ (\hat{p}_{T}^{\ v})_\parallel ~.
\ee

Let us estimate the size of this shift. Since for a non-relativistic boost, $\gamma -1\simeq \frac{1}{2}\beta^2\gamma^2$, the second term in the r.h.s. of (\ref{DeltaETv}) will normally be the dominant one, because it scales with $\beta$ instead of $\beta^2$. However this term is sometimes positive and sometimes negative (depending on the sign of $\beta$), whereas the first term, which scales with $\beta^2$, is always positive. On the other hand, in average, $(\hat{p}_{T}^{\ v})_\parallel\sim\frac{1}{\sqrt{2}}\hat{p}_{T}^{\ v}$. Using eq.(\ref{ETvpole}) we can express  $\hat{p}_{T}^{\ v}$ as a combination of the masses involved in the system, 
\be
\label{pTvpole}
(\left.\hat{p}_T^v)^2\right|_{\rm pole}= \frac{1}{4M_{\chi_2}^2}\left[M_{\chi_2}^2 -M_\chi^2+M_v^2\right]^2 - M_v^2 ~.
\ee
This gets simplified when $M_v^2 \ll M_{\chi_2}^2$, e.g. when the decay occurs via Higgs or $Z-$boson and also in the other cases if we select events with relatively small $M_v$. Therefore, at the pole
\be
\label{pTvpole2}
(\hat{p}_{T}^{\ v})_\parallel\simeq \frac{1}{\sqrt{2}} \frac{1}{2M_{\chi_2}}\left(M_{\chi_2}^2 -M_\chi^2\right) \simeq \frac{1}{\sqrt{2}} \left.\hat{E}_T^v\right|_{\rm pole} ~,
\ee
which substituted back in  eq.(\ref{DeltaETv}) gives 
\be
\label{DeltaETv2}
\Delta\hat{E}_T^v\simeq\left( \frac{1}{2}\beta^2\gamma^2 -\frac{1}{\sqrt{2}} \beta\gamma\right) \hat{E}_T^v  ~. 
\ee
We recall that this is expression is valid for events lying at the pole, which of course are especially interesting as they provide the maximum in the histogram of the signal. Now for a numerical evaluation of $\Delta\hat{E}_T^v$ we need to estimate $\beta\gamma$. In Fig. \ref{fig:pTN2} we show a scatter plot of different SUSY models, showing the correlation between the  transverse momentum of $\chi_2^0$ and its mass. As it can be checked from the fitting function, it is a good approximation to take, in average: 
%It can be empirically checked that typically the transverse momentum of $\chi_2^0$ is in average
%%%%%%%%%%%%%%%%%%%%%%%%%%%%%%%%%%%%%%%%%%%%%%%%%%
\begin{figure}[t]
\centering 
\includegraphics[width=0.45\linewidth]{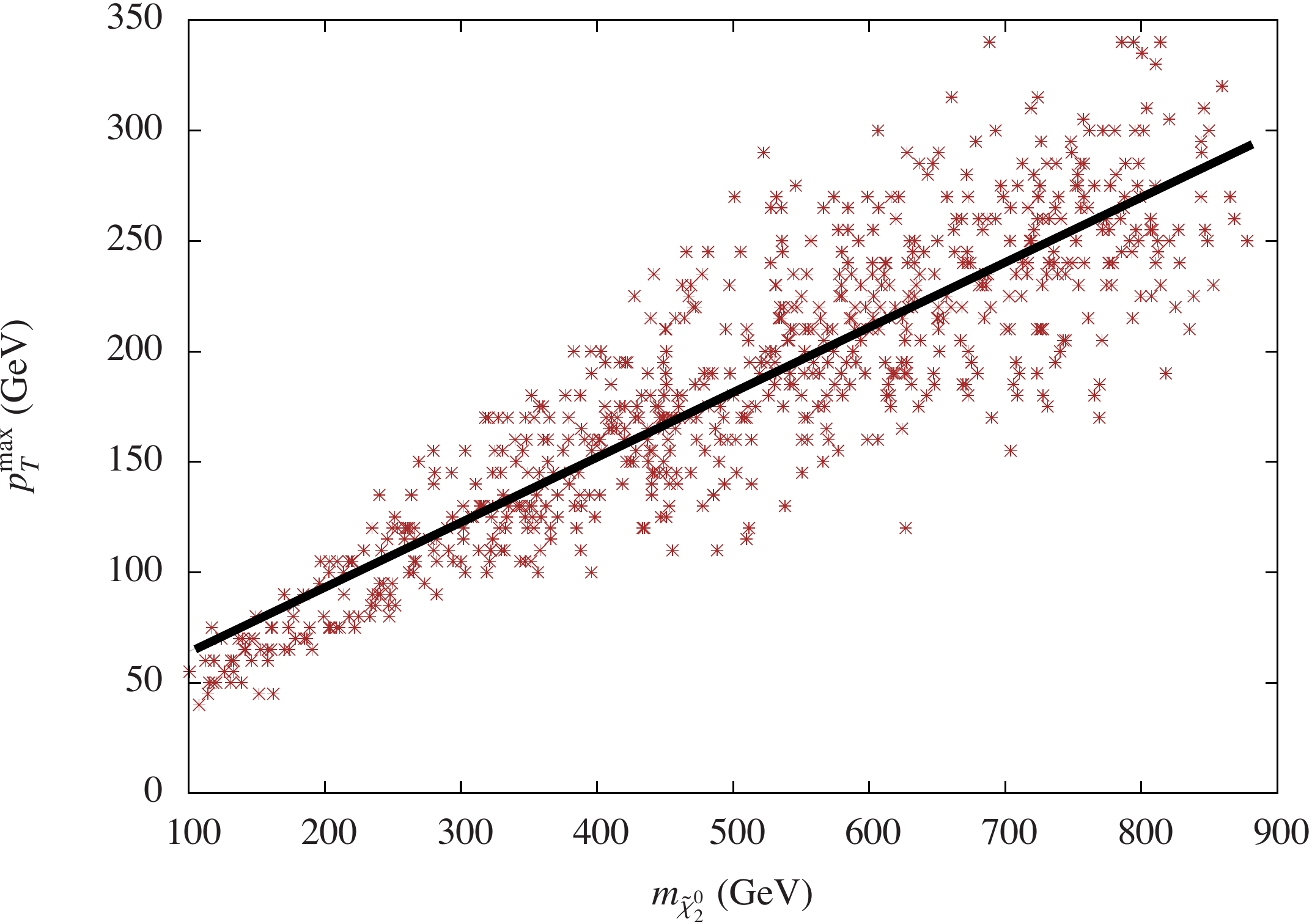}
\includegraphics[width=0.49\linewidth]{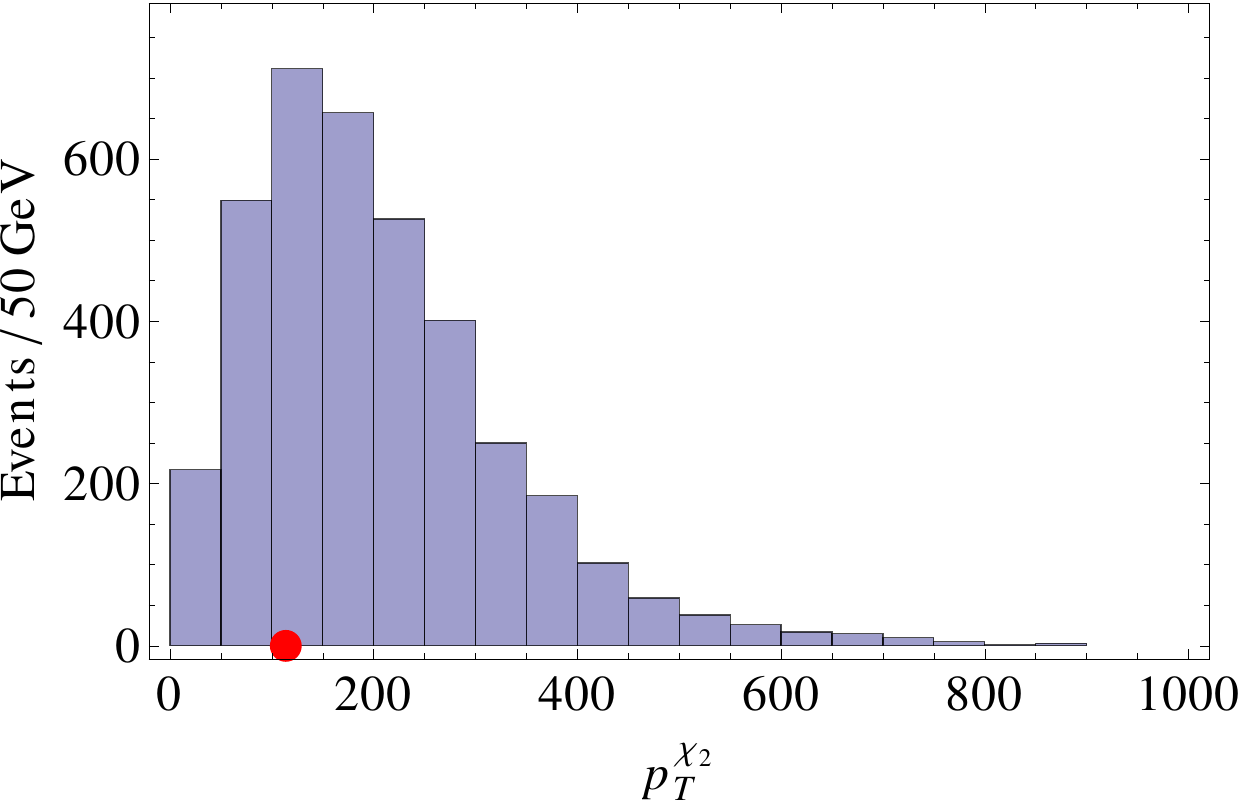}
\caption{{\small Left: Scatter plot of different SUSY models showing the correlation between the $\chi_2^0$'s transverse momentum at the peak of the distribution ($p_T^{\rm max}$) and the $\chi_2^0$ mass. The solid line shows the fitting function $p_T^{\rm max}=aM_{\chi_2}+b$, with $(a,b)=(0.29,35)$. Right: $p_T$ distribution of the $\chi_2$ particle for a particular SUSY model with $M_{\chi_2}=340$ GeV. The red dot shows the $p_T=M_{\chi_2}/3$ value. }}
\label{fig:pTN2}
\end{figure} 
%%%%%%%%%%%%%%%%%%%%%%%%%%%%%%%%%%%%%%%%%%%%%%%%%%
\be
\label{pTtipico}
\left|p_T^{\chi_2}\right| \simeq \frac{1}{3} M_{\chi_2}~.
\ee
%and this value gets slightly smaller as $M_{\chi_2}$ increases. 
This is illustrated in the right plot of Fig.~\ref{fig:pTN2} for a typical example.
Actually this is consistent with expectations. First, the production of $\chi_1^\pm \chi_2^0$ through a $W$  in $S$-channel (see Fig.\ref{fig:C1N2}a) is dominated by the resonance. Even though the $W$ will be normally off-shell, the penalisation for large momenta of $\chi_1^\pm \chi_2^0$  is important, although this reason becomes weaker for $M_W\ll M_{\chi_2}+M_{\chi_1^{\pm}}$. Second, and more importantly, the PDFs penalize large energies. Since the final state particles are colourless (in this case charginos and neutralinos) with a net electric charge, they are essentially produced via quarks/antiquarks (and not gluons). This is significantly affected by the PDFs of the anti-quarks, which tend to prefer lower momenta rather than higher ones. As a consequence of these two effects, the energies of the particles at the final state are expected to be very close to their mass. This is consistent with the relation (\ref{pTtipico}), which implies that the energy of the particle produced differs less than 5\% from its mass. In consequence, from eq.(\ref{gammabeta}, \ref{pTtipico}) typically $\beta\gamma\simeq 1/3$. Then eq. (\ref{DeltaETv2}) becomes
\be
\label{DeltaETv3}
\Delta\hat{E}_T^v=\left( \frac{1}{18}\pm \frac{1}{3\sqrt{2}} \right) \hat{E}_T^v \ ,
\ee
where we have already incorporated the fact that $\beta$ has a random sign. This implies a shift in the range $[-0.2\hat{E}_T^v$, 0.3$\hat{E}_T^v$]  in the visible transverse energy of the pole events when going from CM$\chi$ to LAB. Finally, we have to evaluate how this modifies the position of the peak. Notice that the 
pole events shifted positively (to the right of the histogram), will not produce any new global maximum as they fall 
in a region of $E_T^v$ where there were almost no events. On the other hand the pole events shifted negatively (to the left of the histogram) will populate a region where there were already events. In addition, that region gets also populated by positive shifts of non-pole events, with lower $\hat{E}_T^v$ values. Consequently, we expect a maximum in a value of $E_T^v$ which is approximately 18\% smaller than the value of the pole in CM$\chi$, see eq.(\ref{ETvpole}). 

%{\em On the other hand, for other, more uncommon cases, a $\sim$30\% increase in the maximum of $E_v^T$ could be observed, for the LAB histogram with respect to the CM histogram. This is exemplified in fig.\ref{fig:scatterETv}. }
%%%%%%%%%%%%%%%%%%%%%%%%%%%%%%%%%%%%%%%%%%%%%%%%%%
\begin{figure}[t]
\centering 
\includegraphics[width=0.45\linewidth]{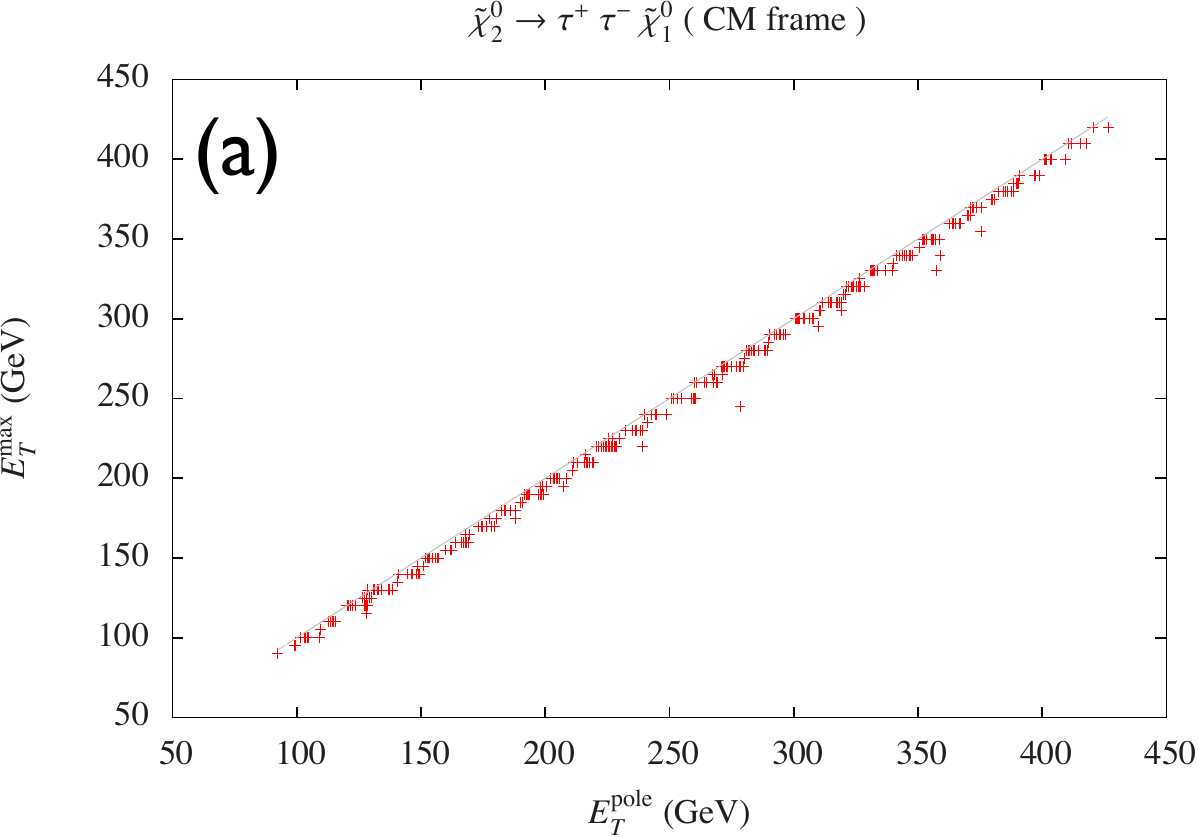}
\includegraphics[width=0.45\linewidth]{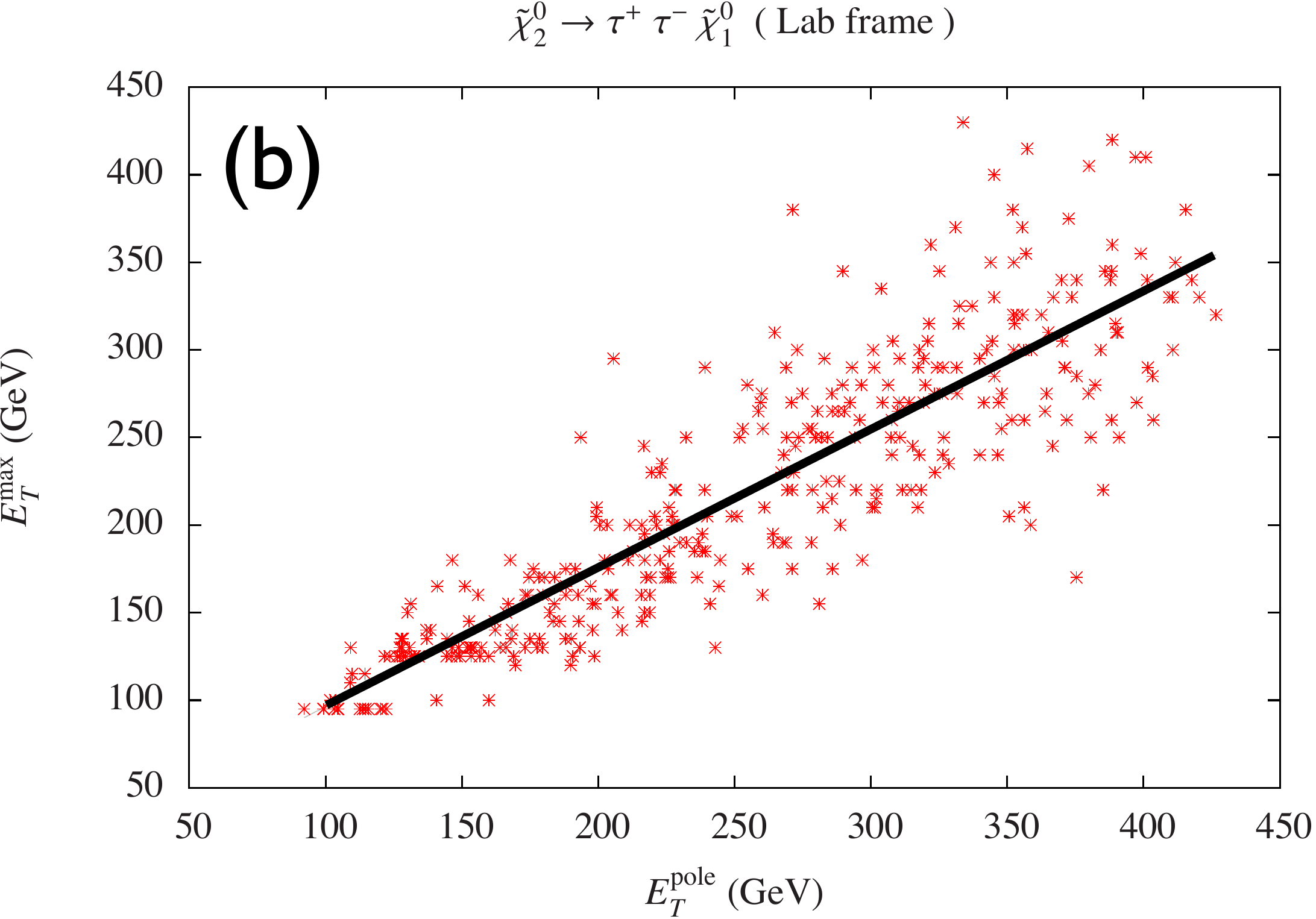}
\hspace{1cm}
\includegraphics[width=0.45\linewidth]{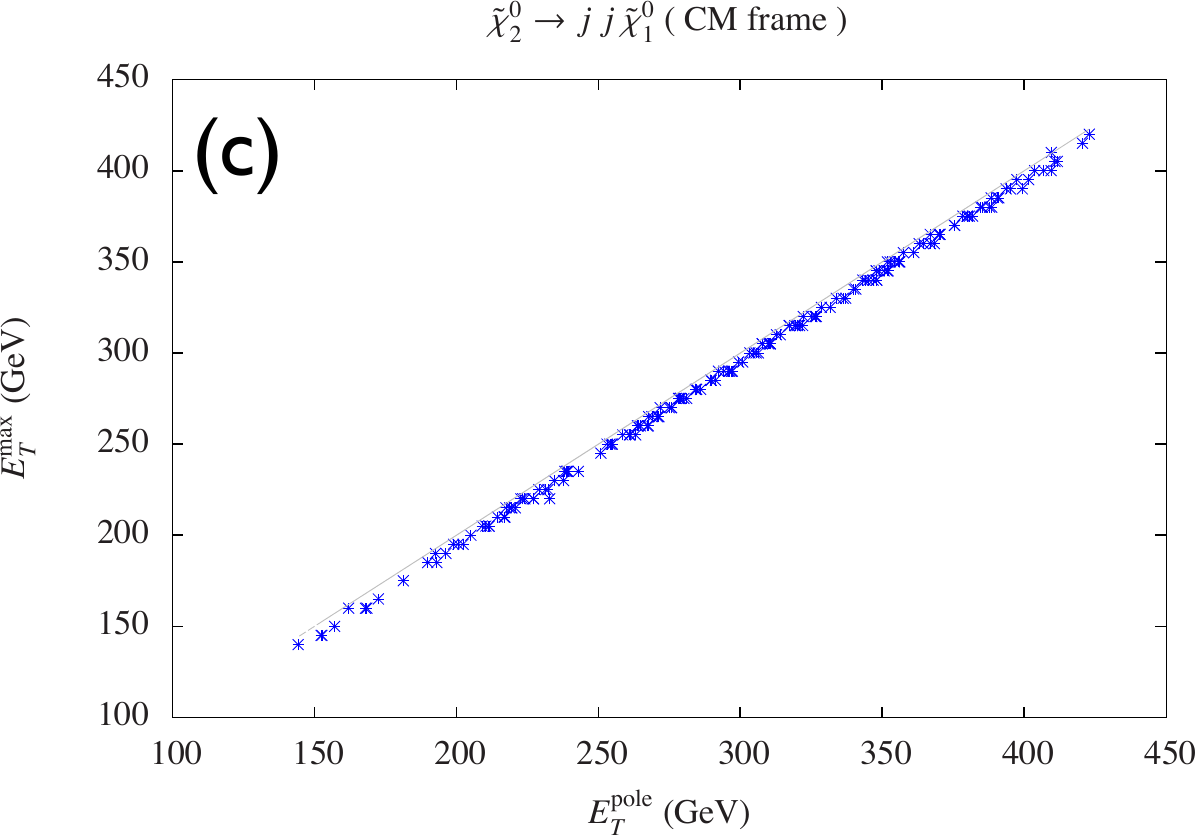}
\includegraphics[width=0.45\linewidth]{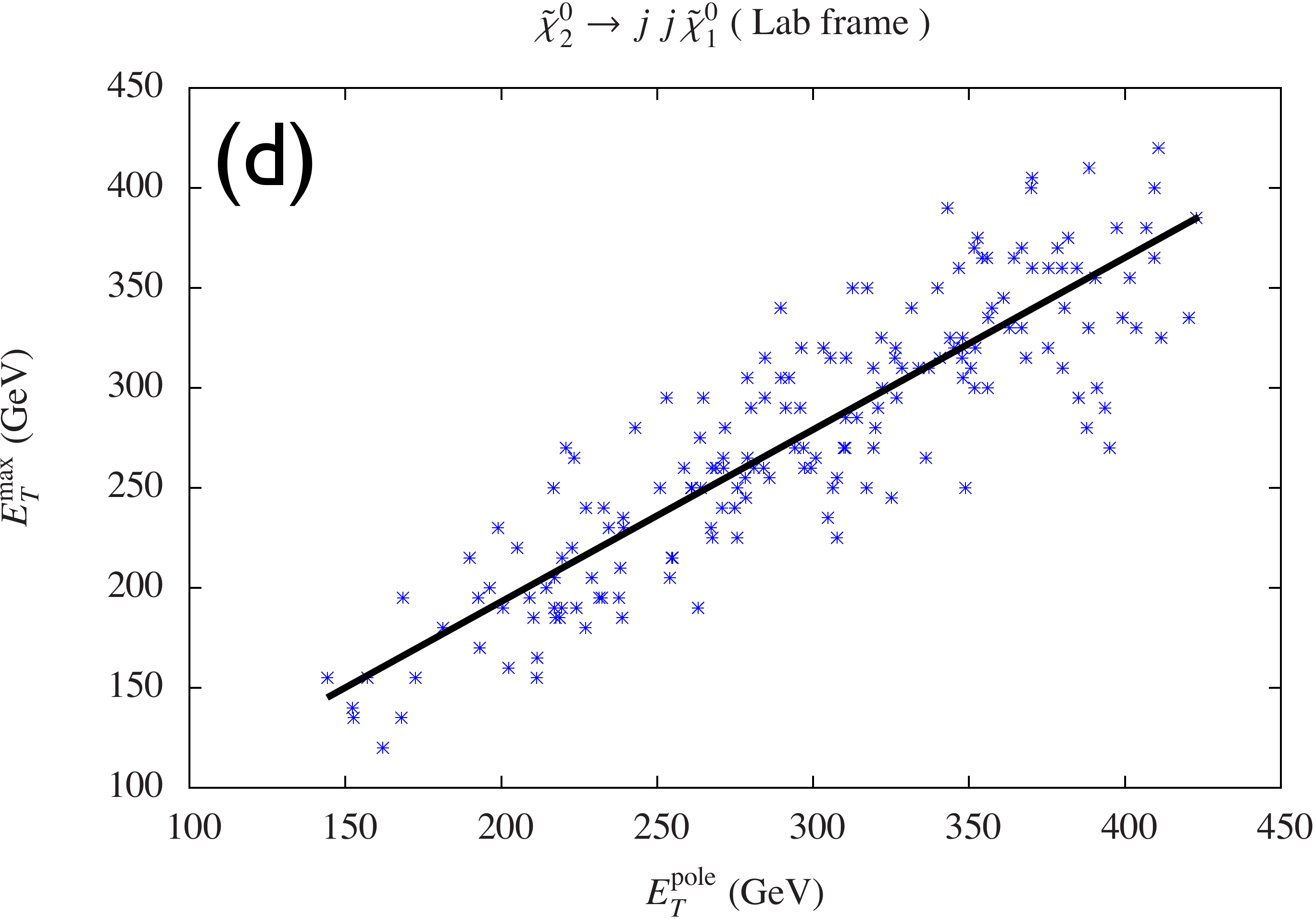}
\caption{{\small Scatter plot of different SUSY models showing the correlation between $E_T^v|_{\rm pole}$ and the peak value of the variable $E_T^v$, reconstructed event by event for each point. This is done for the cases: a) $\tau\bar\tau$ pair in the CM$\chi$ frame; b) $\tau\bar\tau$ pair in the LAB frame; c) $j_bj_b$ pair in the CM$\chi$ frame, and d) $j_bj_b$ pair in the LAB frame. The straight lines show the corresponding fitting functions (see text for details). }}
\label{fig:scatterETv}
\end{figure} 
%%%%%%%%%%%%%%%%%%%%%%%%%%%%%%%%%%%%%%%%%%%%%%%%%%

Fig.\ref{fig:scatterETv} shows scatter plots from a large set of SUSY models, showing the correlation between the theoretical $\hat{E}_T^v|_{\rm pole}$, given by eq.(\ref{ETvpole}), and the actual peak value of the $E_T^v-$histogram, reconstructed for each SUSY model. The upper (lower) plots correspond to histograms of events where the $\chi_2^0$ neutralinos decayed into a $\tau\bar\tau$- ($j_bj_b$-) pairs. We remark again that these plots are shown for illustrative purposes only, so no attempt of realistic identification of the $\tau$ and $b-$jets states is done at this level\footnote{In this sense, a more realistic analysis is performed in the next subsection}. The left (right) plots correspond to histograms in the CM$\chi$ (LAB) system. The CM$\chi$ plots show the perfect correlation of the histogram maximum and the theoretical pole (\ref{ETvpole}). The LAB plots show the two effects expected: a certain spreading and a net shift of the average maximum with respect to the CM$\chi$ prediction. We have fitted the points of the LAB plots with simple linear functions $E_T^{\rm max}=aE_T^{\rm pole}+b$, getting $(a,b) = (0.79\pm0.02,17\pm6)$, 
 $ (0.86\pm0.04,22\pm11)$
for the $\tau\bar\tau$ and $j_bj_b$ cases, respectively.
Note that the results for $\tau\bar\tau$ events are consistent with the previously discussed $\Delta \hat{E}_T^v\simeq -18\%\hat{E}_T^v$ expectation. It is remarkable, however,  that for $j_bj_b$ events, the LAB results are  more symmetrically distributed around the CM$\chi$ prediction. This can be easily understood by recalling that for $j_bj_b$ events the CM$\chi$ histogram does not have an end-point at the pole because of the limited efficiency in the jet reconstruction, see plot c) of Fig.2. In consequence the discussion after eq.(\ref{DeltaETv2}) gets now modified since the pole $j_bj_b-$events shifted positively fall 
in a region of $E_T^v$ where there were already events. In consequence, the limited efficiency in the jet reconstruction funnily makes the position of the maximum more stable than for the leptonic histogram; thus the symmetric spreading of the LAB scatter-plot around the CM$\chi$ prediction.

One of the most interesting features of the $E_T^v$ variable is that it concentrates the signal events around a maximum (determined by the supersymmetric spectrum), which does not happen for the background events (these get concentrated around $M_Z$, see below). This behaviour is very helpful to separate the signal events from the background events, thus improving the signal/background (S/B) ratio in the search of new physics at LHC (even before using this variable to extract information about the SUSY spectrum). It should be noticed here that the background events (typically SM production of $WZ$) behave as if the "neutralino" $\chi_2^0$ had exactly the mass of the $Z$-boson and the "neutralino" $\chi_1^0$ had zero-mass. Thus for the background events the peak in the $E_T^v$ variable lies at $\sim M_Z$.

For some SUSY models, however, the position of the maximum of the signal -determined by eq.(\ref{ETvpole})- can be close to $M_Z$, i.e. the peak of $E_T^v$ for background events. Then, the strategy to detect the existence of new-physics events can be further improved by plotting the $p_T^v$ variable rather than $E_T^v$ (of course, both variables are equivalent since they are unambiguously related through $E_T^v = \sqrt{M_v^2 + (p_T^v)^2}$, where $M_v$ is a measurable quantity). Note that for the background events, the peak in $p_T^v$ is not far from the lower kinematical cut used, while for the signal events is given by eq.(\ref{pTvpole}), and lies normally at some non-trivial value (even if the corresponding $E_T^v$ is close to $M_Z$). In the next section we will see some explicit examples where the use of the $p_T^v$ variable is very convenient to show the existence of new physics in the first place. Once the new physics is detected, the value of $p_T
 ^v$ at the maximum of the signal can be related to the SUSY spectrum through eq.(\ref{pTvpole}) or equivalently through eq.(\ref{ETvpole}).

Next we explore these features in further detail.

\section{Testing the efficiency of $E_T^v$ and $p_T^v$ in concrete SUSY models}
\label{sec:Testing}

As we have described before, the chargino-neutralino pair production can be studied through the use of the visible transverse energy, $E_T^v$, regardless of the way the neutralino $\chi^0_2$ has decayed. So here we only care about initial and final states. 

We will simulate LHC signals with $\sqrt{s}=$14 TeV and luminosity of 100/fb,  using  the package {\tt SUSY-HIT}  \cite{Djouadi:2002ze,Muhlleitner:2003vg,Djouadi:1997yw,Djouadi:2006bz}, as well as {\tt SOFTSUSY}\cite{Allanach:2001kg} for the spectrum calculators, and {\tt MadGraph} {\tt /MadEvent} \cite{Alwall:2011uj} and {\tt PYTHIA}\cite{Sjostrand:2006za} for the event simulation. We focus on events with 3 leptons + missing transverse momentum, and apply the following general cuts
\begin{itemize}
\item the existence of at least two identical, opposite signed leptons  
\item the 3rd hardest lepton having $E_T> 10$ GeV
\item At least an electron (muon) with $E_T>$25 GeV ($p_T>$20 GeV)
\item 3 leptons with $p_T>20$ GeV and $\eta<2.47$ for electrons ($\eta<$2.4 for muons)
\item $E_T^{\rm miss}>$  50 GeV
\item The transverse mass of the 2nd chain $M_T>90$ GeV,
\footnote{Computed with the unpaired lepton, see discussion below. }
\item Jets $p_T < 20$ GeV
\item PGS ATLAS cuts \cite{Alwall:2011uj} ~.
\end{itemize}
The jet reconstruction is performed using the anti-$k_T$ algorithm with $\Delta R=0.4$, whereas the Initial State Radiation is simulated directly from the Matrix Element, with Parton Shower matching implemented by {\tt Madgraph/MadEvent} by making use of MLM methods.

\noindent
We illustrate our strategy by working in the context of the MSSM. Specifically we choose the following SUSY models (defined at the $M_Z$ scale)
\bea
&{\rm Model \ 1:}&\hspace{1cm} M_1 \simeq 99 ~ {\rm GeV}, ~~~~M_{2} \simeq 183~{\rm GeV}, ~~~~\mu\simeq 705~ {\rm GeV}, ~~~~ \tan\beta=10~,\;\;
\nonumber\\
&{\rm Model \ 2:}&\hspace{1cm} M_1 \simeq 47~ {\rm GeV}, ~~~~M_{2} \simeq 244~{\rm GeV}, ~~~~\mu\simeq -515~ {\rm GeV}, ~~~~ \tan\beta=19~,\;\;
\nonumber\\
&{\rm Model \ 3:}&\hspace{1cm} M_1 \simeq 93~ {\rm GeV}, ~~~~M_{2} \simeq 405~{\rm GeV}, ~~~~\mu=-5372~ {\rm GeV}, ~~~~ \tan\beta=50~\;\;
\nonumber\\
&~&
\label{models}
\eea
where $M_1,M_2$ are the bino and wino mass parameters at low-energy, and $\tan\beta \equiv \langle H_u\rangle/\langle H_d\rangle$ is the ratio between the VEVs of the two Higgs doublets. We do not specify the values of gluino and sfermion masses, which are assumed to be heavy (note that we are not imposing gaugino-mass unification). In the three models we study the signal of $\chi_2^0$ decaying through a $Z-$boson, which is the dominant one for all of them (more details below). The dominant background is $WZ$ and $WZ+$jet production in the SM. 

\vspace{0.2cm}
\noindent
The strategy of the analysis for the three models is the following: 
\begin{itemize}
\item We reconstruct the possible invariant masses $M_v$ of a pair of identical leptons, selecting the events where a pair of opposite-sign leptons has an invariant mass close to $M_Z$
\[M_v = M_Z \pm 10\  {\rm GeV} ~.\]  
This pair is identified as a daughter of a $Z$, thus the third lepton, $\ell'$, should come from a $W$. This holds for both the signal and the SM background.

It is important to note that most of the $WZ$ background events are actually removed thanks to the $M_T>90$ GeV requirement in the above list of cuts. The reason is the following. Taking into account that for the events of the background the missing momentum can be identified with the neutrino, one constructs the associated transverse mass, which satisfies the following inequality
\be
\label{MTMW}
(M^{\ell'\nu}_T)^2 = (E_T^{\ell'} + E_T^\nu)^2 - (\vec{p}_T^{\ \ell'} + \vec{p}_T^{\ \nu})^2 \leq M^2_W~,  
\ee	
where $\ell'$ is the charged lepton coming from the $W$ (i.e. the unpaired lepton). Hence, 
identifying ${p}_T^{\rm miss}$ with ${p}_T^{\nu}$ and discarding events with $M^{\ell'\nu}_T<M_W$ would in principle remove the whole $WZ$ background. Although, due to the inefficiency in the reconstruction of the missing piece, some background events may violate in practice the inequality (\ref{MTMW}). This turns out to be an extremely efficient constraint to improve the S/B ratio.

\item With the surviving events, we reconstruct the  $E_T^v$,  $p_T^v$ variables associated to the lepton pair daughter of the $Z$.

\item We simulate the background taking into account only the contribution from $WZ$ and $WZ+$jet production, which is by far the dominant one in the regions of interest (see e.g. \cite{Aad:2012cwa}). We have checked that the results of our simulations are in agreement with \cite{Aad:2012cwa}. 

\item Finally, we construct the $E_T^v$,  $p_T^v$ histograms to show the signal over the background.

\end{itemize}

%\subsection{Numerical results}

Next we expound the results obtained for the three models considered, in a separate way.

\vspace{0.2cm}
\noindent
{\bf{\em Model 1}}

From the initial parameters listed in eq.(\ref{models}), we obtain the relevant supersymmetric spectrum for the analysis, which reads
\be
\label{model1} M_{\chi^{\pm}_1}\simeq 207\  {\rm GeV}, \;\; M_{\chi^{0}_2}\simeq 203\ {\rm GeV},\;\;M_{\chi^{0}_1}\simeq 107\ {\rm GeV}.
\ee
Since the staus are heavy, a priori the preferred decay channel of $\chi_2^0$ is through the lightest Higgs, $m_h=126$ GeV. However, since $M_{\chi_2} - M_\chi < m_h$, this channel is suppressed and the favourite decay-channel turns out to be through an on-shell $Z$. As mentioned above, the main background for this final-state topology is the SM production of a $W/Z$ pair.

%%%%%%%%%%%%%%%%%%%%%%%%%%%%%%%%%%%%%%%%%%%%%%%%%%
\begin{figure}[t]
\centering 
\includegraphics[width=0.5\linewidth]{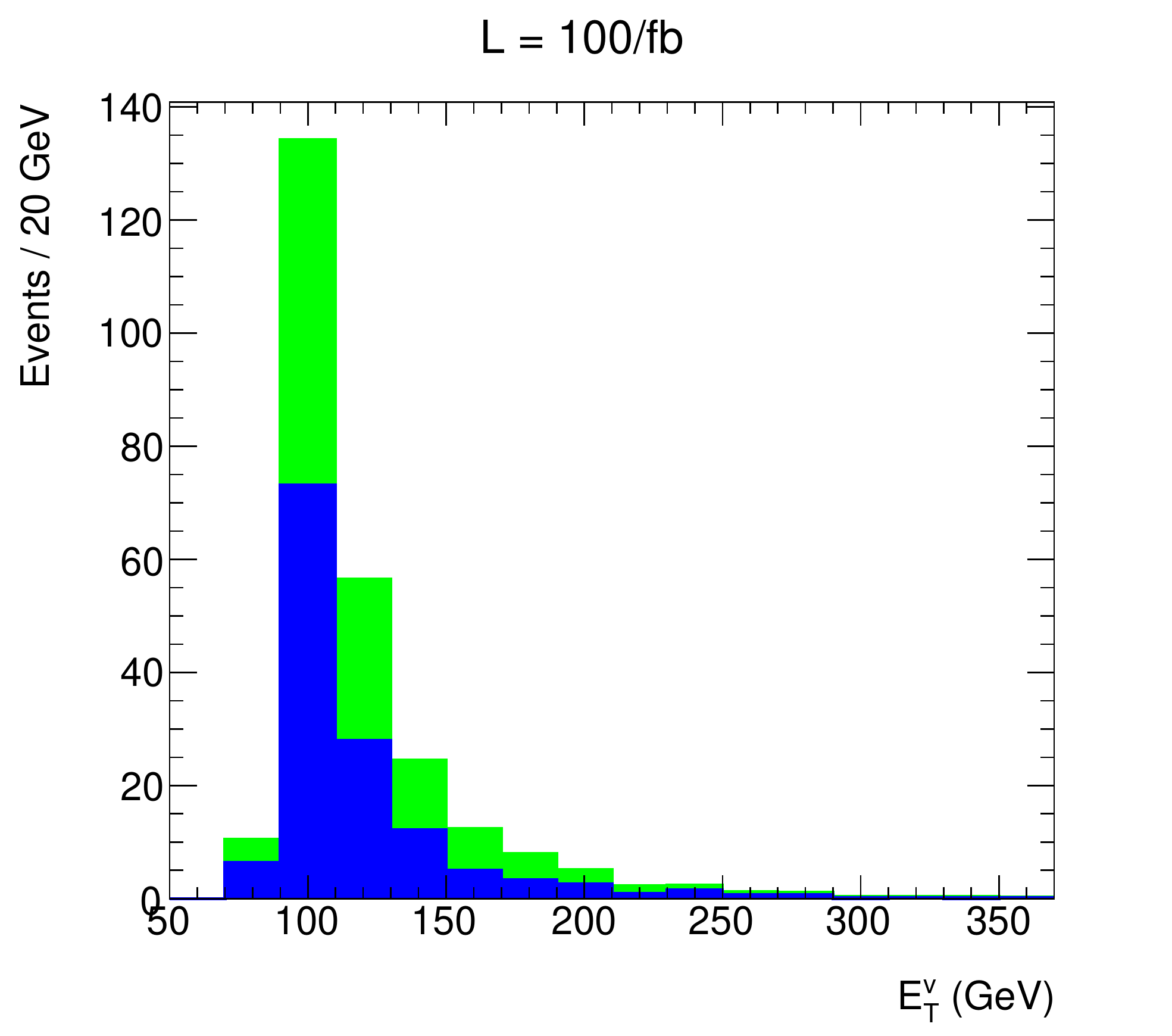}
\caption{ \label{fig:ETv_1}
Histogram (in number of events) of $E_T^v$ for the parameters corresponding to Model 1 (see text), taking into account the signal plus the dominant background (W/Z production). }
\end{figure} 
%%%%%%%%%%%%%%%%%%%%%%%%%%%%%%%%%%%%%%%%%%%%%%%%%%

Fig.\ref{fig:ETv_1} shows the histogram corresponding to the variable $E_T^v$, considering the background plus the signal with a luminosity of 100/fb, which will be reached in the near future. The SUSY signal is clearly visible. The position of  its peak can be used to extract information about the SUSY spectrum in the way discussed in sect. 3. More precisely, for this model the prediction of eq.(\ref{ETvpole}) for the visible transverse energy at the pole of the signal is $\left. \hat{E}_T^v\right|_{\rm pole}\simeq 94$ GeV, which is consistent with the bin of the histogram corresponding to the maximum signal, i.e. $90~ {\rm GeV}-110$ GeV. 

This case illustrates the possibility that the peak of the $E_T^v-$histogram for the signal lies close to the peak for the background, i.e. $M_Z$. Note from Fig. \ref{fig:ETv_1} that both peaks are at the same bin, as expected.

\begin{figure}[t]
\centering 
\includegraphics[width=0.49\linewidth]{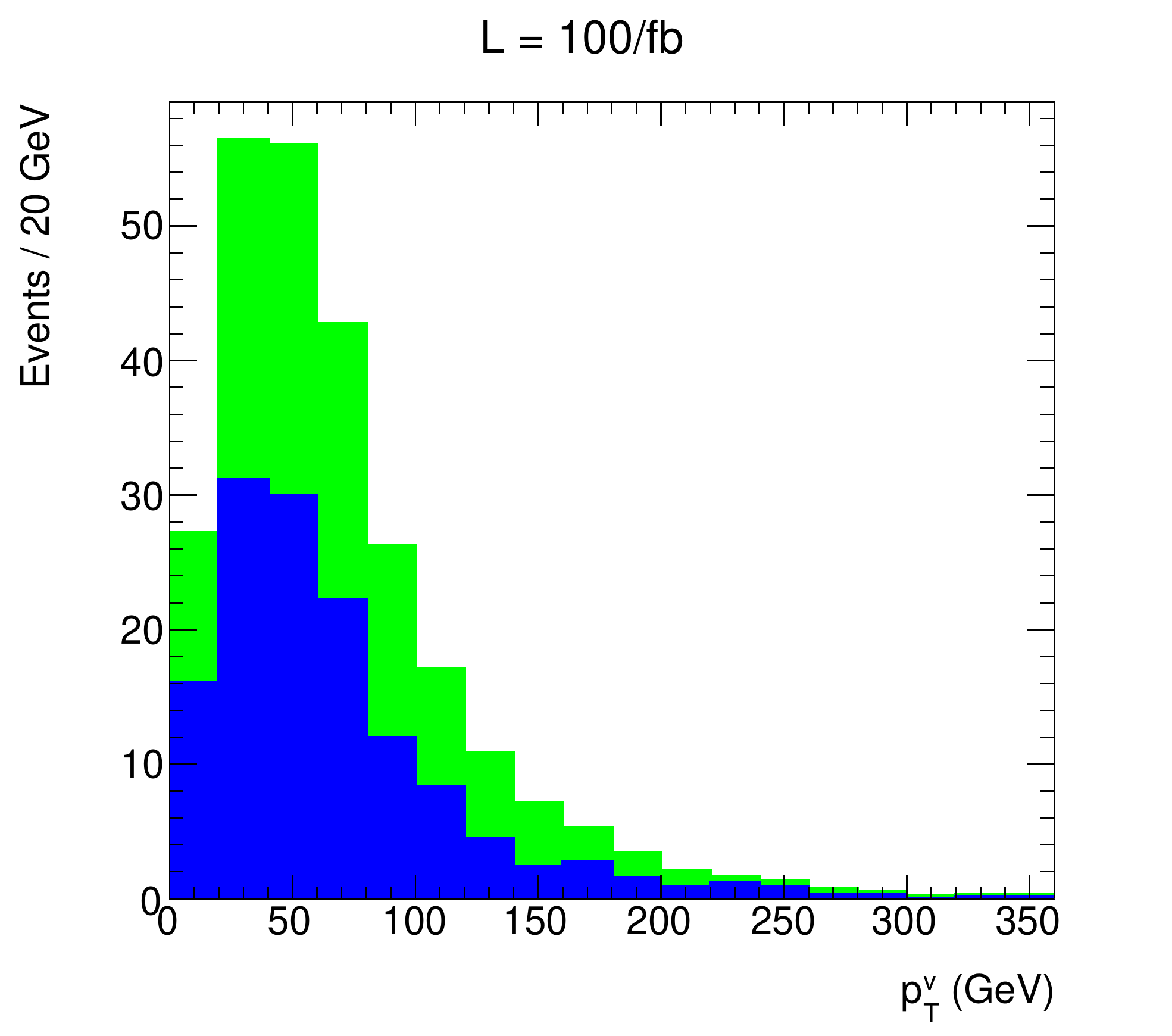}
\includegraphics[width=0.49\linewidth]{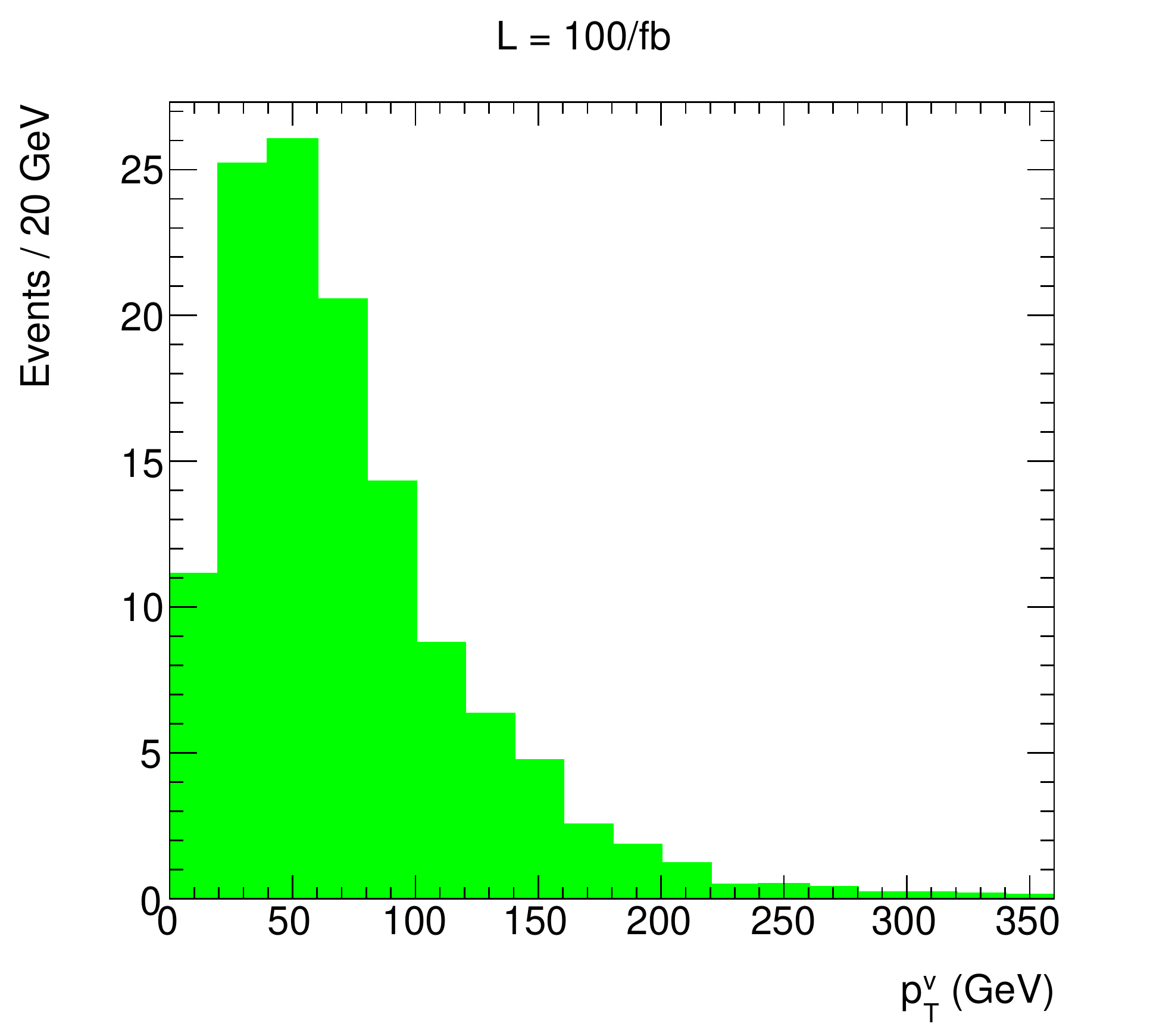}
\includegraphics[width=0.49\linewidth]{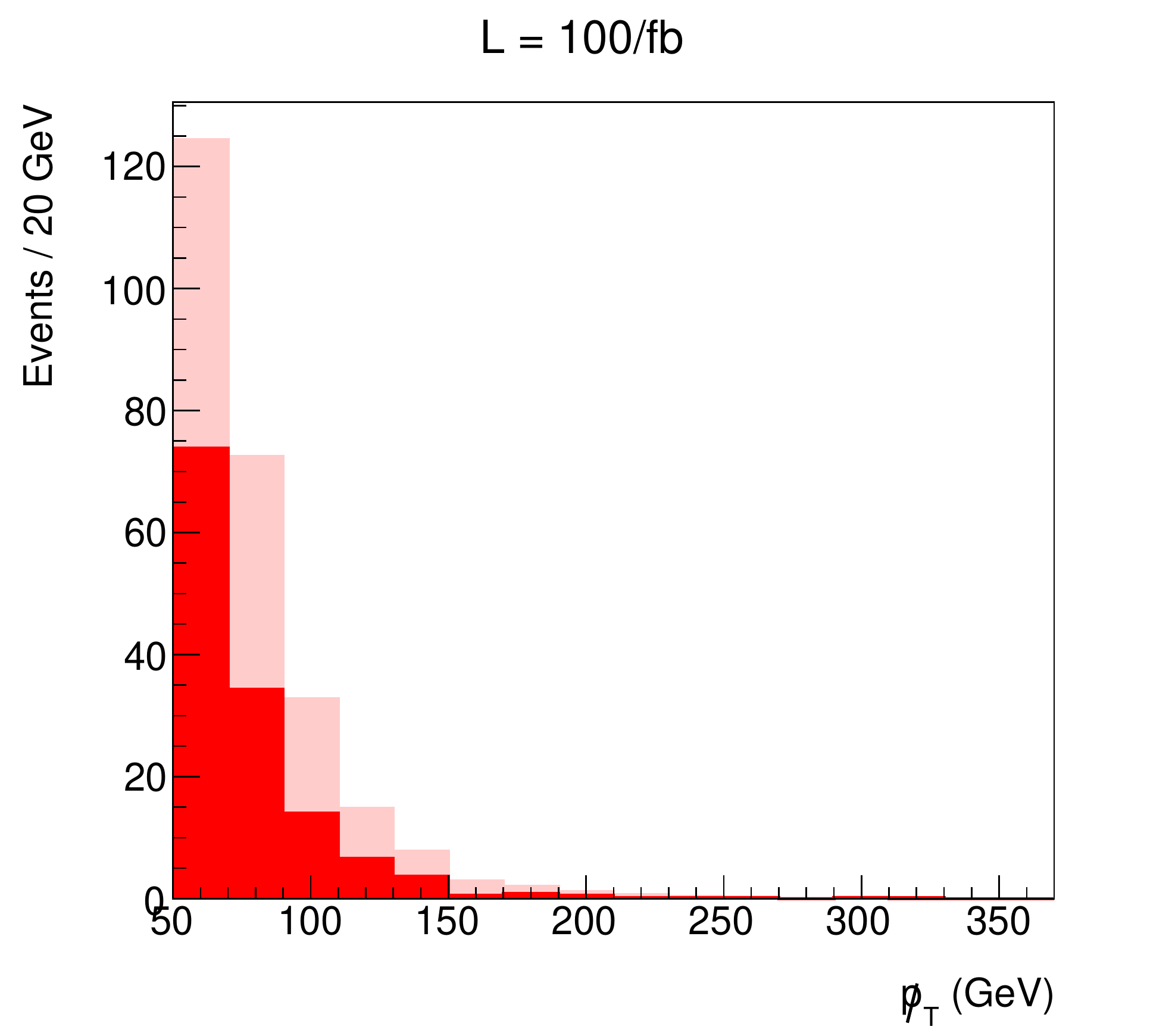}
\includegraphics[width=0.49\linewidth]{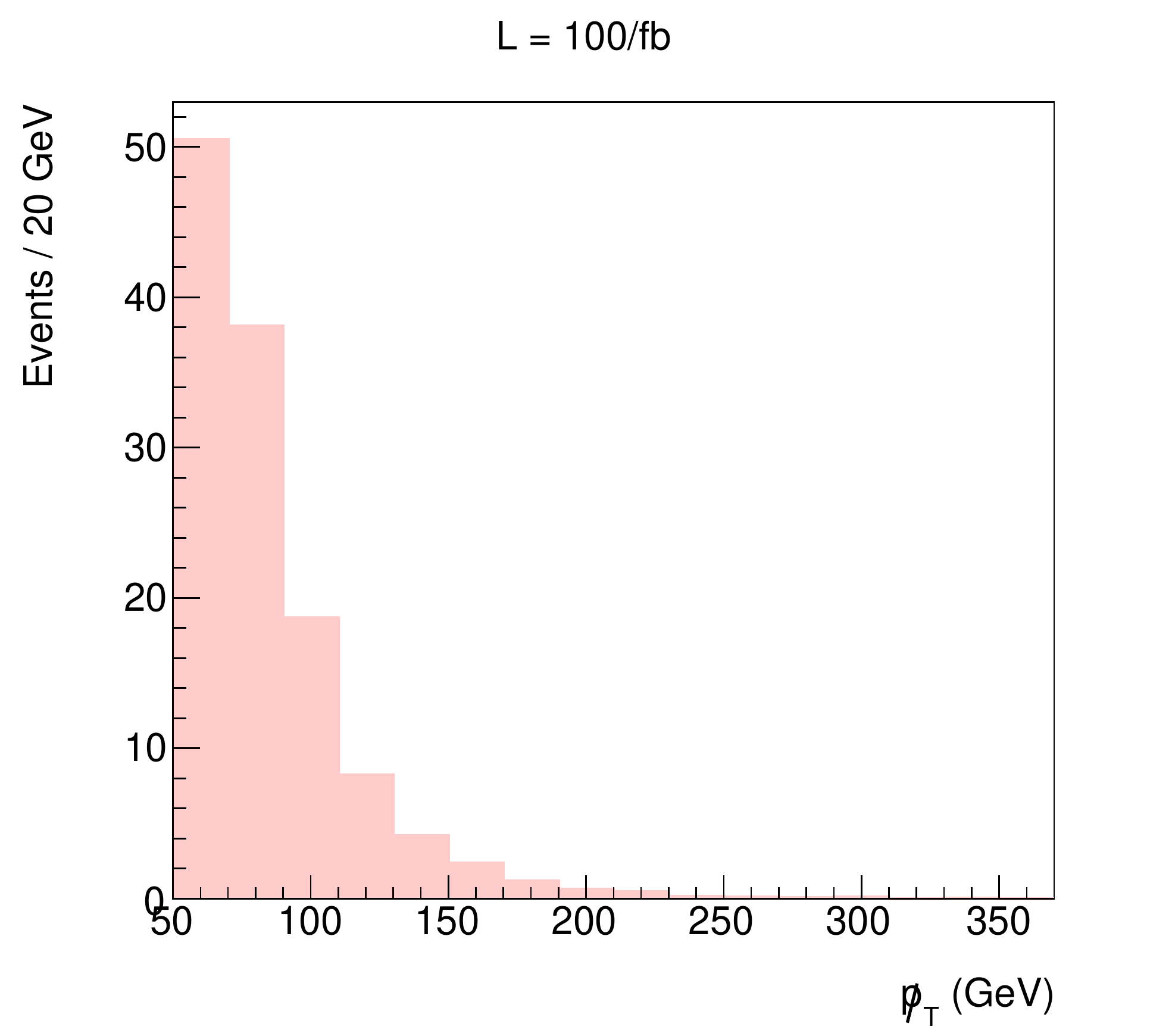}
\label{fig:pTv_1}
\caption{Top- (bottom-) left: the same as in Fig.~5 but now for the $p_T^v$ }
\end{figure}

As discussed at the end of sect. 3, this feature can be improved by plotting $p_T^v$ instead of $E_T^v$. The result is shown in Fig.\ref{fig:pTv_1} (top-left plot). Note that the peak of the signal (the bin centered at 50 GeV) is now displaced with respect to the background one (the bin at 30 GeV), which represents a certain (though admittedly non-dramatic) improving. Fig.\ref{fig:pTv_1} (top-right plot) shows the $p_T^v-$histogram of just the signal events. Of course, it is not something one can realistically obtain from the experiment but we present it in order to show the shape of the signal events, gathered around the (theoretically predicted) maximum.

For the sake of comparison we have presented in Fig.~\ref{fig:pTv_1} (bottom) the analogous plots for the missing momentum variable, $p_T^{\rm miss}$, which is used in the experimental searches of ATLAS and CMS. The $p_T^v$ variable turns out to be slightly better in the S/B ratio, showing that it can be at least as efficient as $p_T^{\rm miss}$ for the initial task of detecting the presence of new physics. Besides, in contrast with  $p_T^{\rm miss}$, the measure of $p_T^v$ is direct and presumably less affected by systematic uncertainties. And, furthermore, once the presence of new physics is recognized, the $p_T^v$ variable provides valuable information about its spectrum since the the maximum of the signal in is related to a defined combination of the supersymmetric masses, eq.~(\ref{pTvpole}).

%%%%%%%%%%%%%%%%%%%%%%%%%%%%%%%%%%%%%%%%%%%%%%%%%%
%%%%%%%%%%%%%%%%%%%%%%%%%%%%%%%%%%%%%%%%%%%%%%%%%%

\newpage
\noindent
{\bf{\em Model 2}}
\newline
\noindent
In this case the relevant supersymmetric spectrum reads
\be
\label{model2} M_{\chi^{\pm}_1}\simeq 273\  {\rm GeV}, \;\; M_{\chi^{0}_2}\simeq 273\ {\rm GeV},\;\;M_{\chi^{0}_1}\simeq 53\ {\rm GeV}.
\ee
%%%%%%%%%%%%%%%%%%%%%%%%%%%%%%%%%%%%%%%%%%%%%%%%%%
\begin{figure}[h]
\centering 
\includegraphics[width=0.5\linewidth]{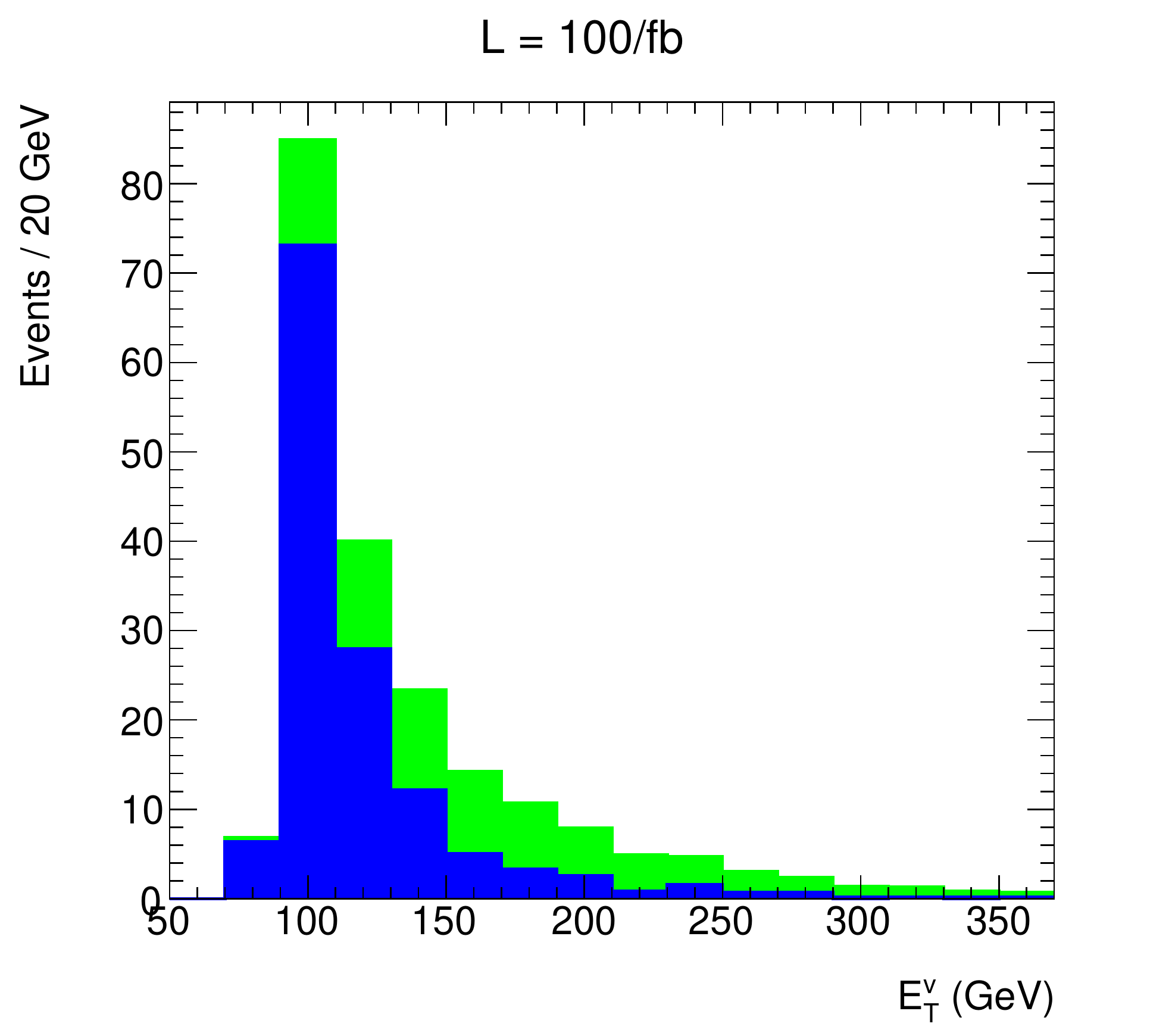}
\caption{The same as Fig. 5, but here for Model 2 (see text). }
\label{fig:ETv_2}
\end{figure} 
%%%%%%%%%%%%%%%%%%%%%%%%%%%%%%%%%%%%%%%%%%%%%%%%%%
%%%%%%%%%%%%%%%%%%%%%%%%%%%%%%%%%%%%%%%%%%%%%%%%%%
\begin{figure}[ht!]
\centering 
\includegraphics[width=0.49\linewidth]{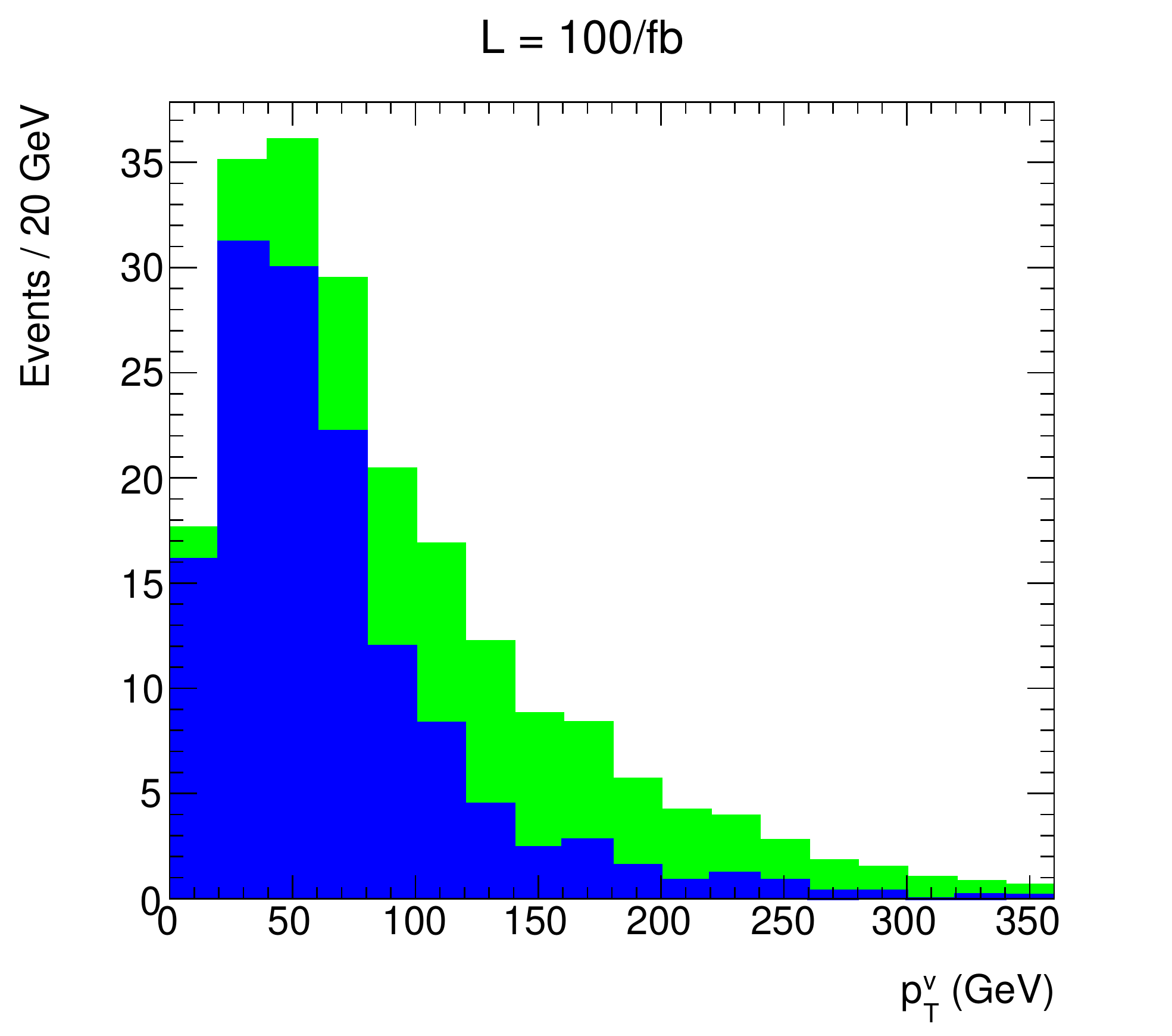}
\includegraphics[width=0.49\linewidth]{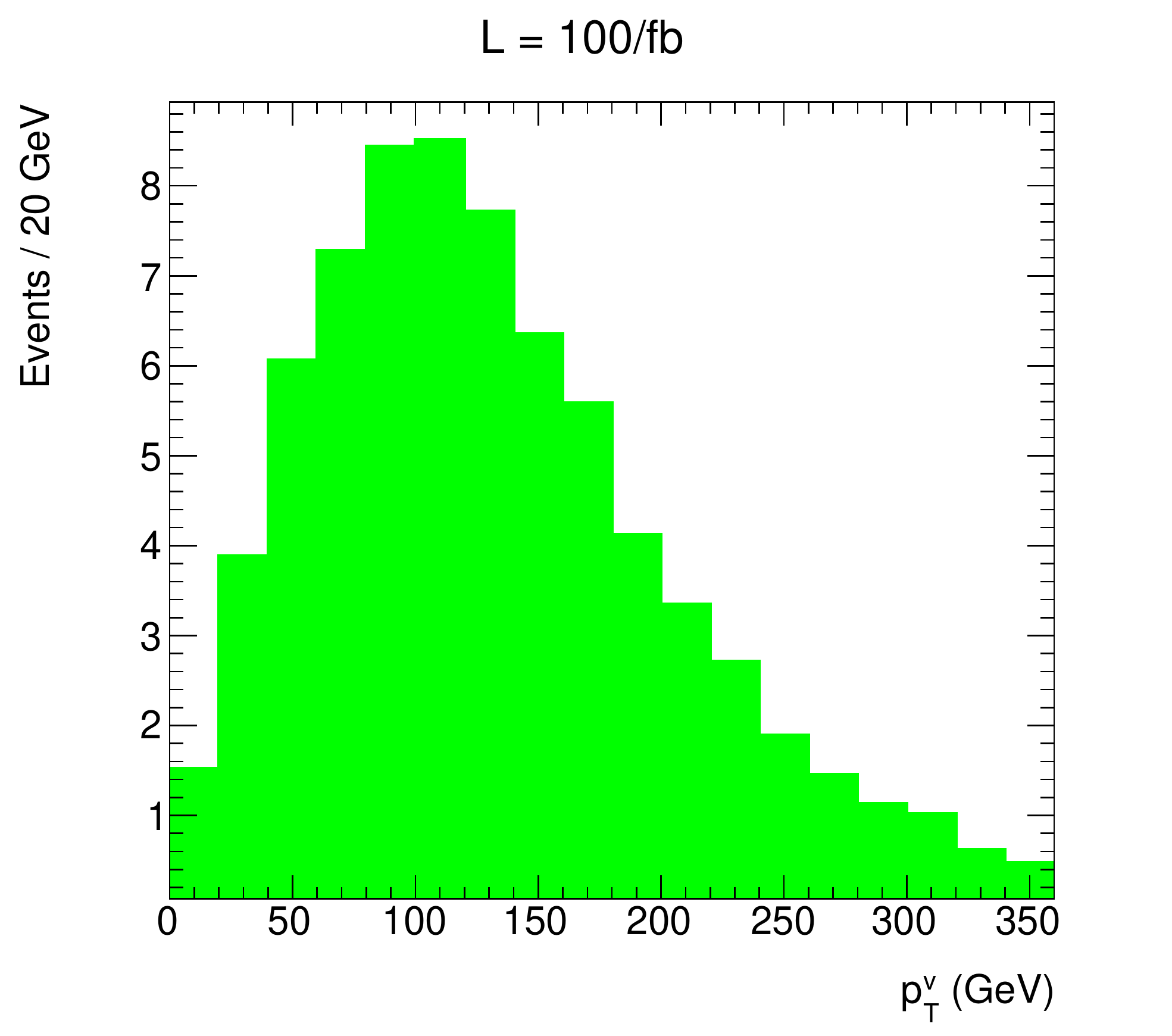}
\includegraphics[width=0.49\linewidth]{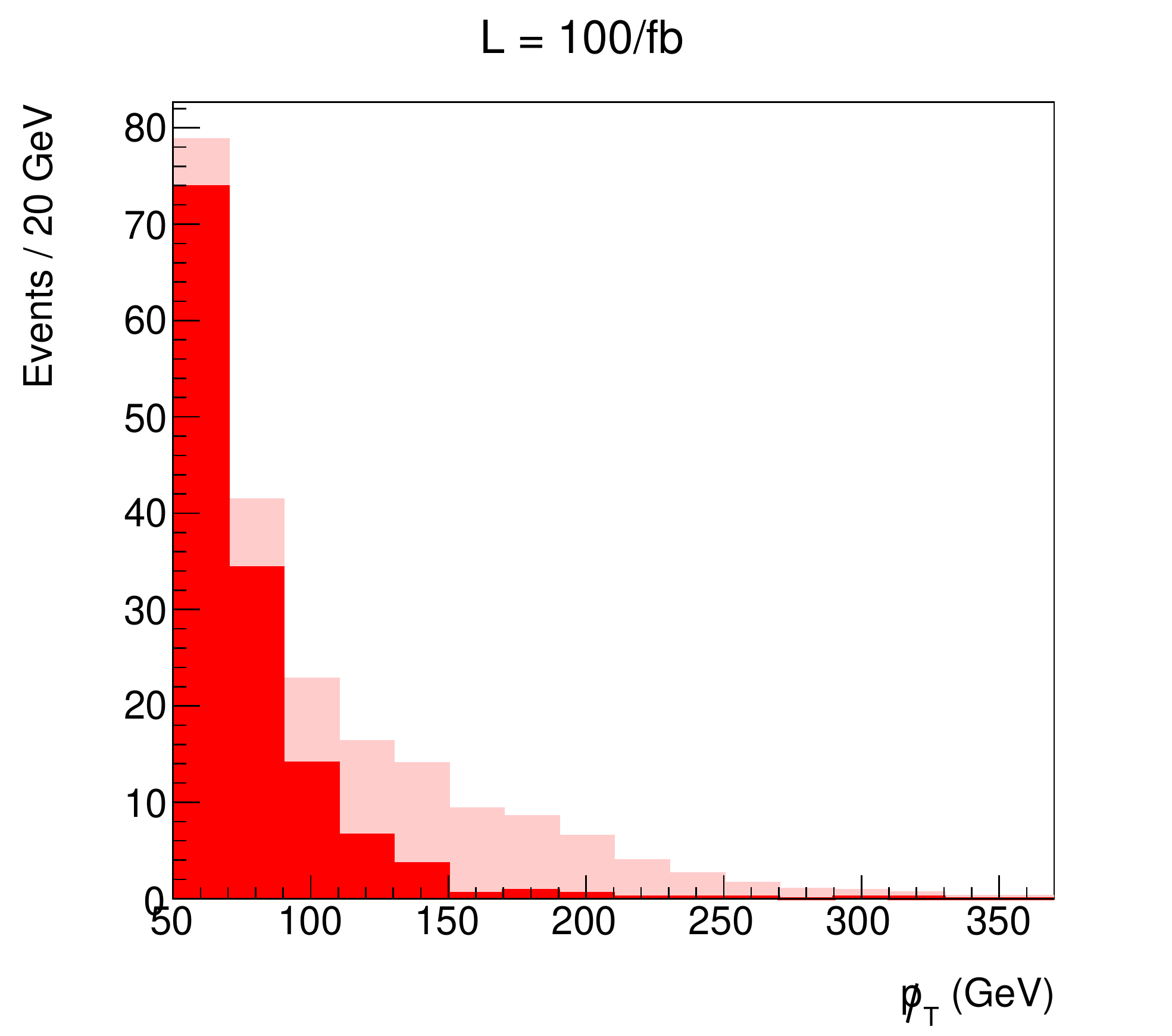}
\includegraphics[width=0.49\linewidth]{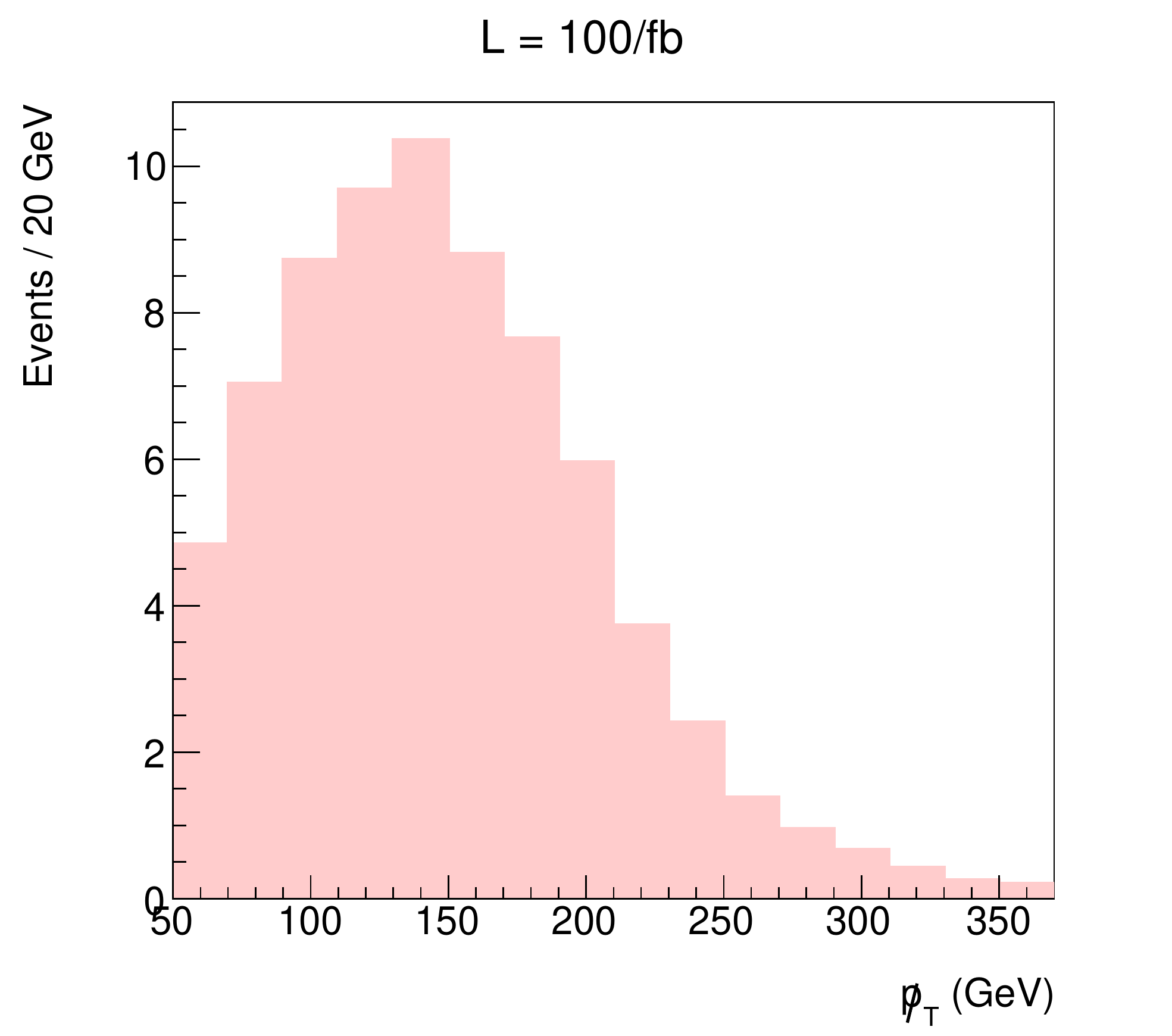}
\caption{The same as Fig.~6, but here for Model 2 (see text). }
\label{fig:pTv_2}
\end{figure} 
%%%%%%%%%%%%%%%%%%%%%%%%%%%%%%%%%%%%%%%%%%%%%%%%%%

%
$\chi_2^0$ and $\chi_1^0$ are almost pure bino and wino respectively.  Now the decay of $\chi_2^0$ through an on-shell Higgs is kinematically allowed, so one would expect it to be the preferred decay channel. However, the branching ratio of the decays through a Higgs or through a $Z-$boson are 39\% and 61\% respectively. The reason is the following. Due to the relatively high value of $\tan\beta$, the $\chi^0_2$ and $\chi^0_1$ neutralinos are essentially gauginos with a small $\tilde H^0_d$ component and a {\em very} small $\tilde H^0_u$ component. On the other hand, the physical Higgs-boson is essentially $H^0_u$ for the same reason. Then the $\chi^0_2$ decay through a Higgs occurs thanks to those very small $\tilde H^0_u$ components and gets suppressed, while the decay through a $Z-$boson may occur also through the not-so-small $\tilde H^0_d$ components of the neutralinos.

 %has to do essentially with the $\tilde H^0_d$ and $\tilde H^0_u$ components of the neutralinos. Given the relatively high value of $\tan\beta$ for this model, it is probable that the produced Higgs would be actually $H^0_u$. However for this model the $\tilde H^0_u$ components of $\chi^0_2$ and $\chi^0_1$ are suppressed by a factor ${\cal O}(0.1)$  with respect to the $\tilde H^0_d$ components. This causes the $\tilde W^0-\tilde H^0_u-H^0_u$ (or $\tilde B-\tilde H^0_u-H^0_u$) couplings to be suppressed, whereas the $\tilde H^0_d-\tilde H^0_d-Z$ aren't.

Fig. \ref{fig:ETv_2} shows the histogram corresponding to the variable $E_T^v$, containing the background plus the signal, again with a luminosity of 100/fb. The SUSY signal is of course weaker than in Model 1, since the supersymmetric masses are larger. On the other hand, it is nice that now the peak of the signal does not coincide with the peak of the background. In this case eq.(\ref{ETvpole}) gives $\left. \hat{E}_T^v\right|_{\rm pole}\simeq 146$ GeV, consistent with what the signal peak visible at Fig. \ref{fig:ETv_2}. 

As for Model 1, in this case the separation between the signal and background peaks is more efficiently achieved using the variable $p_T^v$ rather than $E_T^v$, as it is shown in Fig.\ref{fig:pTv_2} (top plots). We show also the analogous plots for {{{}}} $p_T^{\rm miss}$. Again the behavior is slightly better for $p_T^v$, in particular in the region of higher statistics.

\vspace{0.2cm}
\noindent
{\bf{\em Model 3}}

Finally, for Model 3 the spectrum reads
\be
\label{model3} M_{\chi^{\pm}_1}\simeq 405.3\  {\rm GeV}, \;\; M_{\chi^{0}_2}\simeq 405.3\ {\rm GeV},\;\;M_{\chi^{0}_1}\simeq 94\ {\rm GeV}.
\ee

Again here $\chi^0_2$ and $\chi^0_1$ are almost purely wino and bino respectively, with very small Higgsino components of ${\cal O}(\lesssim 10^{-2})$. Also in this case the neutralinos contain much more $\tilde H^0_d$ than $\tilde H^0_u$, leading to branching ratios BR$(\chi^0_2\to \chi^0_1 h)$ and BR$(\chi^0_2\to \chi^0_1 Z)$ to be 4\% and 96\% respectively, for similar reasons as before. Since now $\tan\beta$ is much larger the Higgs channel is even more suppressed.

\begin{figure}[ht!]
\centering 
\includegraphics[width=0.5\linewidth]{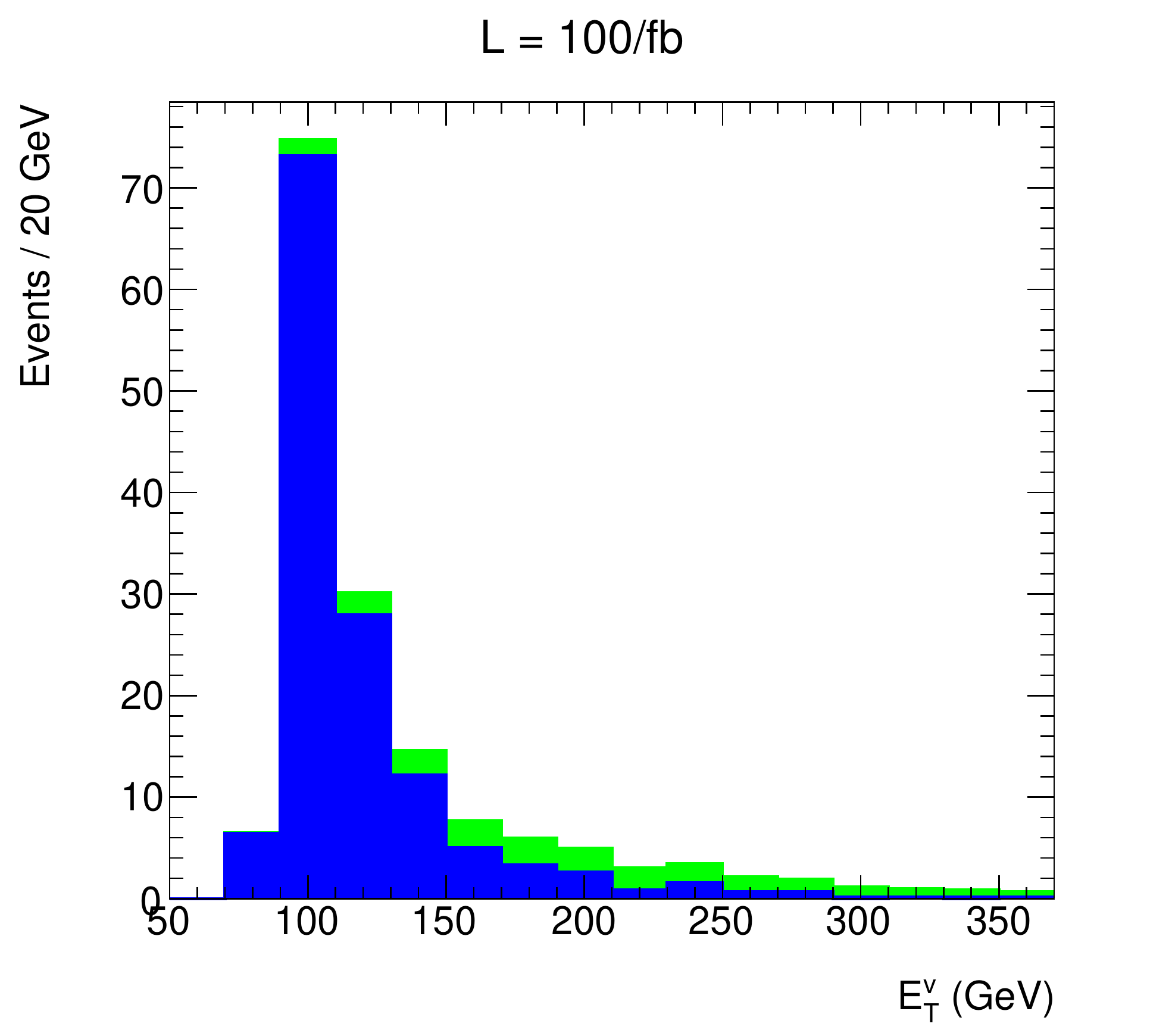}
\caption{The same as Figs.~5,  7, but here for Model 3 (see text).  }
\label{fig:ETv_3}
\end{figure}

\begin{figure}[ht!]
\centering 
\includegraphics[width=0.49\linewidth]{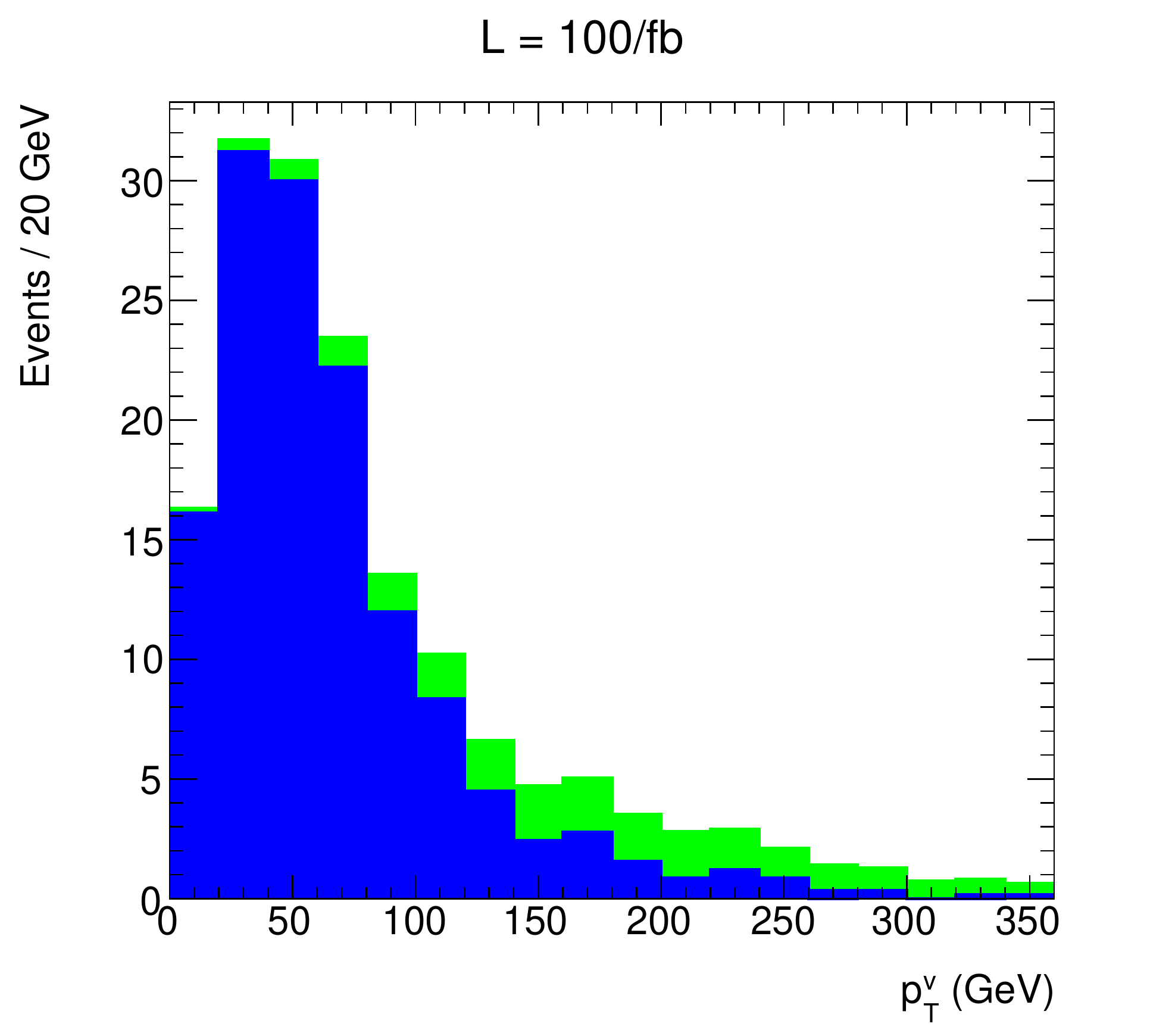}
\includegraphics[width=0.49\linewidth]{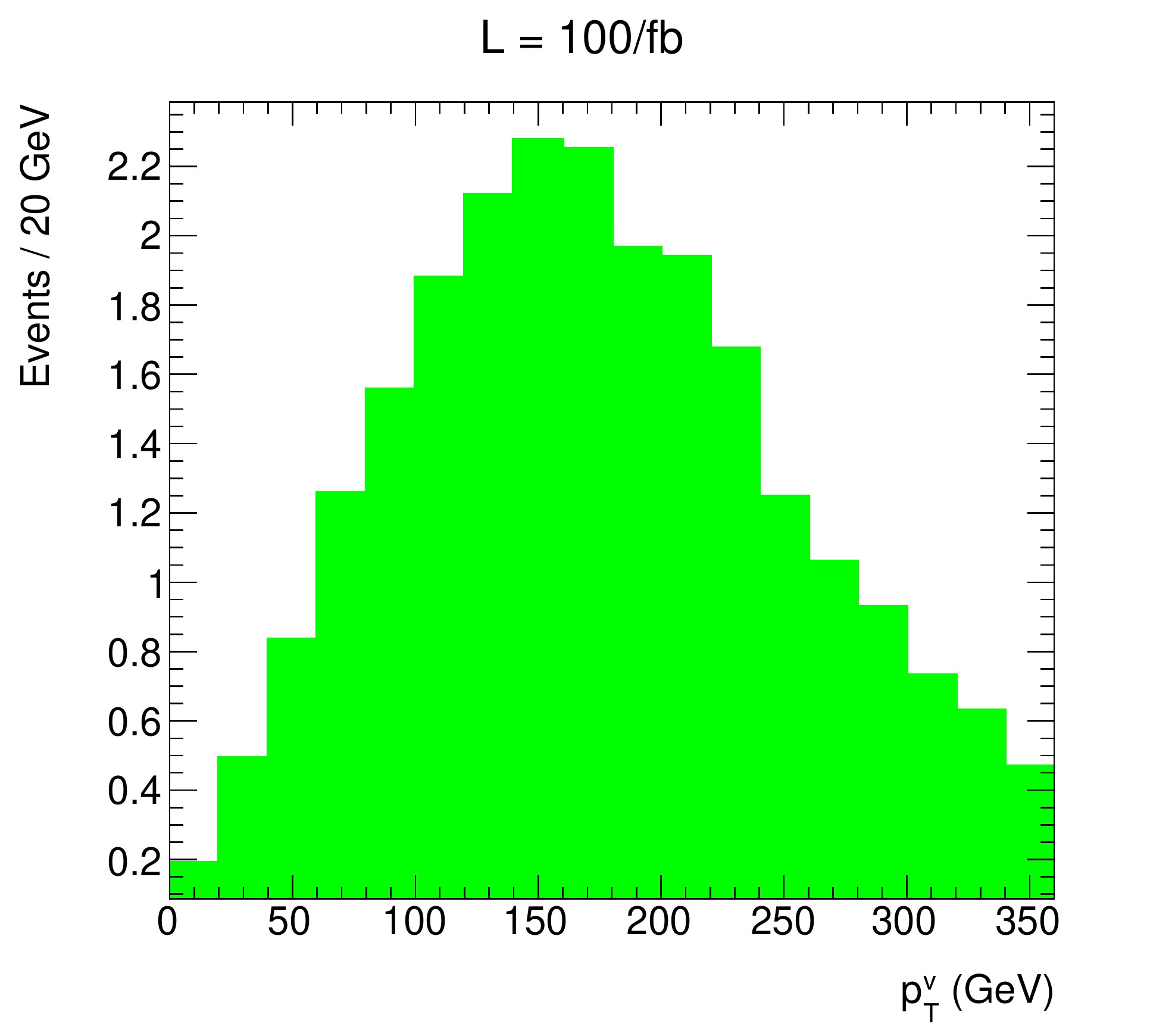}
\includegraphics[width=0.49\linewidth]{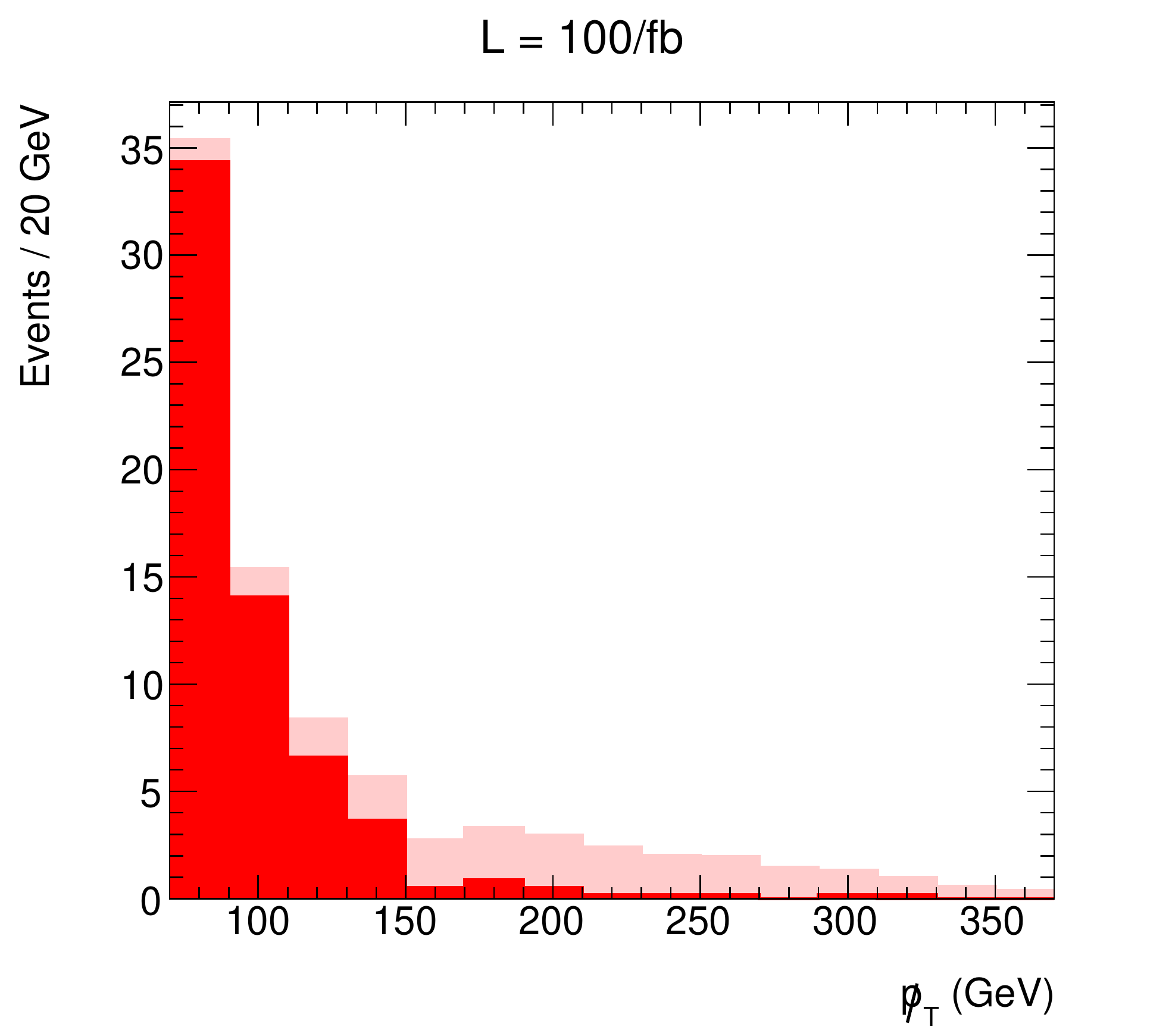}
\includegraphics[width=0.49\linewidth]{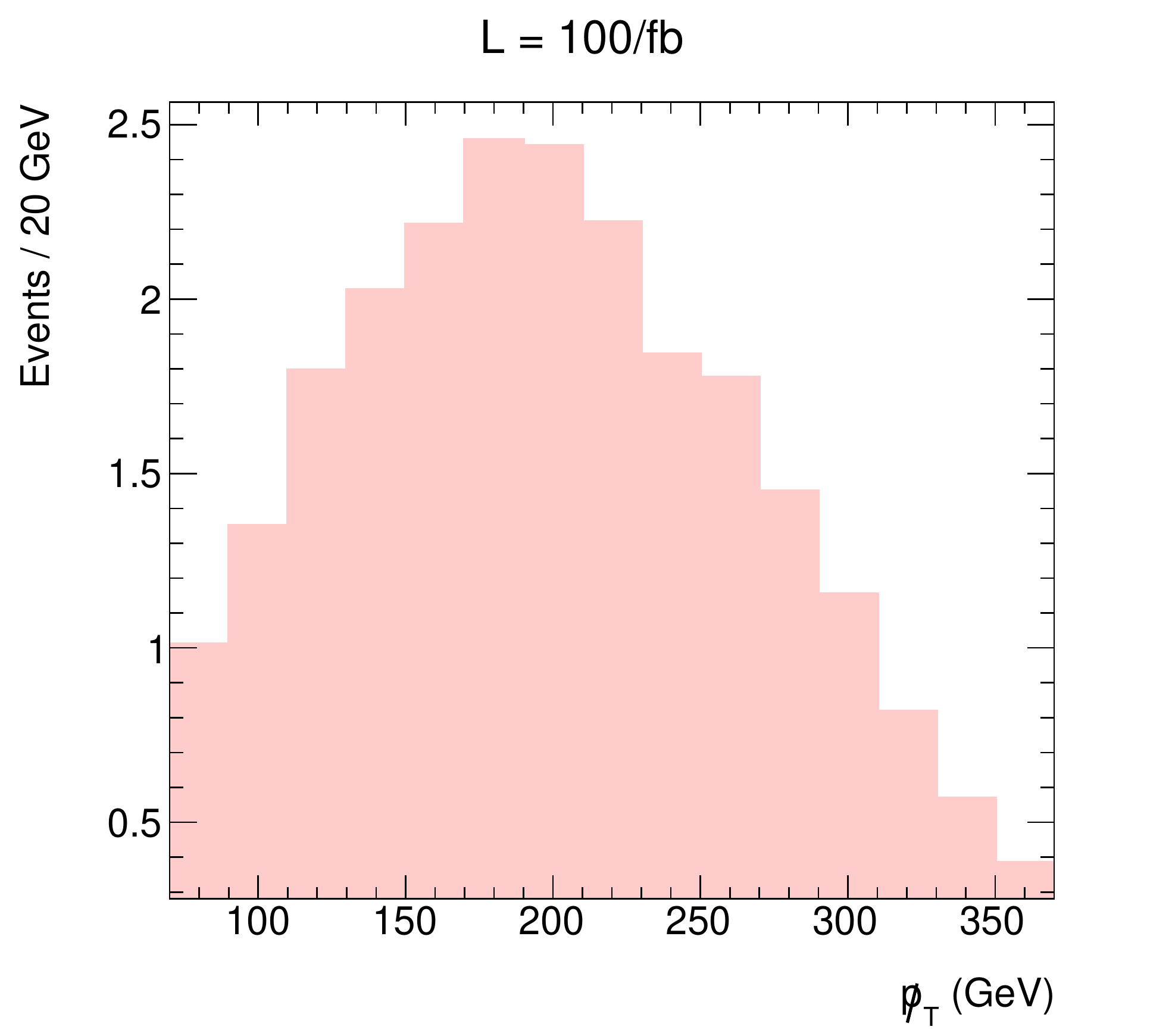}
\caption{The same as Fig.~6, 8, but here for Model 3 (see text). }
\label{fig:pTv_3}
\end{figure}

Fig.~\ref{fig:ETv_3} and Fig.~ \ref{fig:pTv_3} show the histograms in the $E_T^v$ and the $p_T^v, p_T^{\rm miss}$ variables respectively. They show similar features as for Model 2. The difference of course is that the signal is now quite small due the large supersymmetric masses. Still it is visible, especially for the $p_T^v$ variable. Since the S/B ratio is high ($\sim 2$) around the $p_T^v$-peak of the signal, the later would be visible with larger luminosities (say 300/fb).

%%%%%%%%%%%%%%%%%%%%%%%%%%%%%%%%%%%%%%%%%%%%%%%%%%
%%%%%%%%%%%%%%%%%%%%%%%%%%%%%%%%%%%%%%%%%%%%%%%%%%
%%%%%%%%%%%%%%%%%%%%%%%%%%%%%%%%%%%%%%%%%%%%%%%%%%
%%%%%%%%%%%%%%%%%%%%%%%%%%%%%%%%%%%%%%%%%%%%%%%%%%

\section{Conclusions}
\label{sec:Concl}

The study of chargino-neutralino pair production at the LHC finds nowadays an unprecedented motivation, since the detection of those particles could well represent an opportunity of discovering light supersymmetric states at the LHC. Different strategies have been used so far by the experimental groups ATLAS and CMS to explore this process. They have been mainly focused on the analysis of 3-leptons plus missing-energy final states using standard kinematical variables, such as $p^{\rm miss}_T$ and the invariant mass $m_{\ell^+\ell^-}$. 

In this work, we have presented a new -purely kinematical- method based on the variable $E_T^v= \sqrt{M_v^2 + ({p}^v_T)^2}$, i.e.  the transverse energy of the visible products coming from the neutralino (typically two leptons or two b-jets), which presents some very useful features. First of all, the histogram in $E_T^v$ (in the frame where the decaying neutralino is at rest) has a pole. This translates into a peak in the actual experimental distribution. Consequently, this is a very robust feature against poor statistics. In addition, this concentration of signal-events around the maximum does not occur for the background-events. In this sense, it is highly advantageous to use, instead of $E_T^v$, the equivalent variable $p_T^v$ (transverse momentum of the visible decaying products). This optimizes the S/B ratio, showing a (slightly) better performance than the usual $p_T^{\rm miss}$ variable. Furthermore, the main merit of the $E_T^v$ or $p_T^v$ variables is that the peak 
 in the histogram is correlated in a well-defined way to a precise combination of SUSY masses, given by
\be
\label{concl}
\left.{E}_T^v\right|_{\rm pole}= \frac{1}{2 M_{\chi_2}}\left[M_{\chi_2}^2 -M_{\chi_1}^2+M_v^2\right]\ ,
\ee
\be
\label{concl2l}
\left. ({p}_T^v)^2\right|_{\rm pole}= \left. ({E}_T^v)^2\right|_{\rm pole} -M_v^2~.
\ee
where $M_{\chi_2}$, $M_{\chi_1}$ and $M_v$ are the masses of the decaying neutralino $\chi_2^0$, the lightest neutralino (LSP) and the visible system to which $\chi_2^0$ decays to. 

Of course, when passing from the CM$\chi$ to the LAB system the maximum becomes less sharp and the position of the maximum is slightly shifted. However, these effects are not dramatic and, besides, can be well estimated in a semi-analytical way and kept under control.

We have illustrated these facts performing a realistic analysis of particular SUSY models, where the $\chi_2^0$ neutralino decays mainly through a $Z$ boson. We have focussed in events where the final state consists of 3 leptons plus missing energy, showing that the SUSY signal in the $p_T^v$ variable could be very well detected at the  LHC running with 14 TeV and a luminosity of 100/fb, which is attainable in the near future.

Let us finally stress that the same method can be applied to any possible decay channel of $\chi^0_2$: through squark, sleptons or a Higgs. The latter case is actually the most frequent one for MSSM models. Then, the most probable final state contains 2 b-jets, 1 lepton (or jet) and missing energy. The $h\to b\bar b$ channel has the additional advantage of having large branching ratios compared to the $Z\to\ell\bar\ell$ case. On the other hand, the topology $j_bj_b\ell$ is much less clean and difficult to reconstruct properly. Thus it requires a separate study which is out of the scope of the present paper and will be the subject of a future research work.

\section*{Acknowledgements}  

We are grateful to Krzysztof Rolbiecki and Paul de Jong for very useful discussions. This work has been partially supported by the MICINN, Spain, under contract
FPA2010-17747; Consolider-Ingenio PAU CSD2007-00060, CPAN CSD2007-00042. We
thank as well the Comunidad de Madrid through Proyecto HEPHACOS S2009/ESP-1473
and the European Commission under contract PITN-GA-2009-237920. M. E. Cabrera
acknowledges the financial support of the CSIC through a predoctoral research
grant (JAEPre 07 00020); and the ERC ``WIMPs Kairos - the moment of truth for
wimp dark matter'' (P.I. Gianfranco Bertone). The work of A. Casas has been partially supported by the MICINN, Spain, under contract FPA2010-17747 and the Consolider-Ingenio PAU CSD2007-00060, CPAN CSD2007-00042. We thank as well the Comunidad de Madrid through Proyecto HEPHACOS S2009/ESP-1473 and the European Commission under contract PITN-GA- 2009-237920.

\bibliography{ccm}{}
\bibliographystyle{JHEP}

\end{document}